\documentclass[10pt,twocolumn,a4paper,fleqn]{article}

\usepackage[square,super,numbers]{natbib}
\usepackage{balance}
\newcommand{\myemail}{{\let\thefootnote\relax\footnote{$\star$
    \texttt{Lukasz.Bratek@pk.edu.pl}}}}
\newcommand{\mytitle}[1]{\begin{quotation}{\bf\huge\noindent #1}
    \end{quotation}}
\newcommand{\myabstract}[1]{\begin{quotation}\noindent{\bf Abstract.}{
    \small #1}\end{quotation}}
\newcommand{\mykeywords}[1]{\begin{quotation}\noindent{\bf Keywords:}{
    \small #1}\end{quotation}}    

\usepackage{url}
\usepackage[utf8]{inputenc}
\usepackage[T1]{fontenc}
\usepackage{amsmath,amssymb}
\usepackage{array}
\usepackage[scriptsize,bf]{caption}
\usepackage{float}
\usepackage{placeins}
\usepackage{xfrac,upgreek}
\usepackage[]{color}
\usepackage{multirow}
\usepackage{microtype}
\usepackage[integrals]{wasysym}
\usepackage{graphicx}
\usepackage{ulem}
\usepackage{hyphenat}
\usepackage{enumitem}
\usepackage[cal=boondoxo]{mathalfa}
\usepackage{xspace} 

\definecolor{myred}{RGB}{150,30,30} 
\definecolor{mygreen}{RGB}{30,150,30} 
\definecolor{mygrey}{RGB}{125,125,125} 
\definecolor{myblue}{RGB}{30,30,150}

\newcommand{\ud}[1]{\mathrm{d}{#1}}
\newcommand{\br}[1]{\left(#1\right)}
\newcommand{\sq}[1]{\left[#1\right]}
\newcommand{\bag}{\mathcal{B}}
\newcommand{\bagref}{\mathcal{B}_{\rm rf}}
\newcommand{\magref}{\mathcal{H}_{\rm rf}}

\newcommand{\Epsilon}{\mathrm{E}}

\newcommand{\Chi}{\mathrm{X}}

\newcommand{\erg}{\mathrm{erg}}
\newcommand{\m}{\mathrm{m}}

\newcommand{\cm}{{\rm cm}}
\newcommand{\fm}{{\rm fm}}
\newcommand{\km}{{\rm km}}

\newcommand{\gauss}{{\rm Gs}}
\newcommand{\Gs}{\gauss}
\newcommand{\MeV}{{\rm MeV}}
\newcommand{\MeVcc}{{\MeV\!/\!{\rm c}^2}}
\newcommand{\GeV}{{\rm GeV}}
\newcommand{\Msun}{{\rm M}_{\odot}}
\newcommand{\gram}{{\rm g}}
\newcommand{\tabref}[1]{Tab.\ref{#1}}
\newcommand{\figref}[1]{Fig.\ref{#1}}
\newcommand{\secref}[1]{Sec.\ref{#1}}
\renewcommand{\eqref}[1]{Eq.\ref{#1}}
\newcommand{\mref}{{\rm m}_{\rm r}}
\newcommand{\epsref}{{\rm p}_{\rm r}}
\newcommand{\nref}{{\rm n}_{\rm r}}
\newcommand{\sat}{_{\sigma}}
\newcommand{\mCL}{m_{150}}
\newcommand{\mycal}[1]{#1}

\newcommand{\repl}[2]{{\color{mygrey}{\sout{#1}}}{\color{myblue}{#2}}}
\newcommand{\add}[1]{{\color{mygreen}{#1}}\xspace}

\newcommand{\rem}[1]{{\color{mygrey}{\sout{#1}}}}

\begin{document}
\twocolumn[\begin{@twocolumnfalse}

\bigskip\bigskip\bigskip

\mytitle{Phenomenological Scaling Relations for SQM Stars with a Massive s-Quark 
in Gravitationally Strong Magnetic Fields under the Spherical Symmetry Approximation}

\medskip

\begin{center}
{\large {\L}ukasz Bratek$^{1,*}$, Joanna Ja{\l}ocha$^{1}$, Marek Kutschera$^{2}$}
\\
\medskip
\begin{tabular}{@{}l}
{\small $^1$Department of Physics, Cracow University of Technology,  
Podchor\k{a}{\.z}ych 1, PL-30084 Krak{\'o}w, Poland}\\
{\small $^2$Jagiellonian University, Institute for Theoretical Physics,  Łojasiewicza 11, PL-30348 Krak\'{o}w, Poland}
\end{tabular}\\
\bigskip 
\texttt{Preprint version.\\
The improved and final version was published in \emph{Phys. Rev. D} \textbf{110}, 083041 (2024),\\
\url{https://journals.aps.org/prd/abstract/10.1103/PhysRevD.110.083041}}
\end{center}

\myabstract{
Scaling relations with the bag constant parameter are investigated for Strange Quark Matter (SQM) stars in the presence of gravitationally strong 
magnetic fields minimally coupled with matter, considering both massless and massive strange quark scenarios. Assuming a simple model for such 
coupling under the approximation of spherical symmetry, a phenomenological scaling formula for the maximum mass of stars is derived as 
a function of the surface magnetic field and the strange quark mass. This formula is applicable for all formally admissible values of the free 
parameters, and strict scaling with the bag constant is retained only for a vanishing strange quark mass.
As a byproduct of this study, the mathematical structure of the equation of state for the relativistic SQM model with a massive strange quark 
is revisited without approximations. It is observed that the contribution of the electron term to the energy density cannot be neglected in the 
theoretical limit of large strange quark masses. Consequently, the maximum mass of SQM stars increases with sufficiently large strange quark mass, 
contrasting with the behavior observed for low strange quark masses.
}

\mykeywords{Unified Astronomy Thesaurus concepts: Neutron stars (1108); Magnetic stars (995); Gravitation (661); Degenerate
matter (367); SQM stars, strange quark matter, equation of state with massive s-quark -- full analysis}

\renewcommand{\thefootnote}{\fnsymbol{footnote}}  
\bigskip
\end{@twocolumnfalse}]
\renewcommand{\thefootnote}{\fnsymbol{footnote}}  
\footnotetext[1]{\texttt{Lukasz.Bratek@pk.edu.pl}}
\setcounter{footnote}{0}                  
\renewcommand{\thefootnote}{\arabic{footnote}}  


\section{Introduction}

This paper examines a static compact star model with a gravitationally strong magnetic field under the approximation of spherical symmetry. 
{Focus is made on} {m}agnetic fields strong enough to exert a noticeable gravitational effect on the star 
structure{,} {while neglecting a further complication 
from altering the equation of stellar material by the magnetic field.}
Specifically, the study addresses Strange Quark Matter (SQM) stars with a massive s-quark, continuing the work presented in \cite{bratek2023}, 
which investigated a symmetric SQM star model with a magnetic field. The emphasis here is on scaling relations with massive s-quark 
and the influence of {strong} magnetic fields on these relations. 

Since the introduction of concepts by \citet{bodmer1971} and \citet{witten1984}, the hypothesis of quark stars has been intensively considered. 
The equation of state for quark matter is straightforward, both in its symmetric form and when considering a non-zero s-quark mass. 
The scaling relations for quantities such as the maximum mass, the corresponding radius, and the mass-radius relation itself, showing how these 
quantities vary with changes in the bag constant $\bag$ and the s-quark mass $m_s$, as well as their interrelations, 
have been studied by \citet{zdunik2000}.

The current study adds to these investigations, emphasizing the connection of the scaling relations with the symmetries of the differential equations of structure. Introducing the s-quark mass makes the scaling relations approximate both with and without external fields. This is because the mass introduces a length scale, allowing comparisons with dimensional quantities of the original model.
In contrast, a strict scaling with $\bag$ is characteristic of a certain group of equations of state 
with a power law dependence on the baryon density. Adding a magnetic field introduces further 
perturbations to the scaling formulas, even with {the suitable} form of coupling between matter and the magnetic field 
{that preserves
the scaling symmetry in the case of symmetric SQM\cite{bratek2023}}. This is due  to the magnetic energy density, which can be compared with the bag constant of the same dimension:
$$\mathcal{H}^2/\mathcal{B},$$ where $\mathcal{H}$ is a scalar describing the magnetic field strength. Despite this, 
{as will be shown,} a strict scaling relation can be regained if the s-quark mass is appropriately scaled so that a dimensionless 
 parameter $b$, involving $\bag$ and $m_s$ through the ratio $$b\propto \bag/m_s^4,$$ remains constant. This makes sufficient to describe the model at some reference bag constant $\bagref$ only. From the properties of the equation of state, it follows that there is a lower limit $b_m$ on the parameter $b$ (and hence on $\bag$ at a specific value of the s-quark mass, or vice versa). 
The presence of such a critical value means that the model cannot be simplified further.

The influence of strong or ultra-strong magnetic fields on the structure of stars is crucial to consider.
Magnetic fields in neutron stars typically range from $10^{12}$ to $10^{13}\Gs$~\cite{taylor1993}, with values reaching 
up to $10^{14}$--$10^{16}\Gs$ in some cases, including $8{\times}10^{14}\Gs$~\cite{kouveliotou1998}, $10^{15}\,\Gs$~\cite{Paczynski1992}, $3{\times}10^{15}\Gs$~\cite{thompson1993} or ultra-strong dipole fields potentially reaching up to $3{\times}10^{17}\Gs$ for millisecond rotational period pulsars \cite{duncan1992}.  Observationally determined surface magnetic fields are in the range of $10^{14}$--$10^{15}\Gs$~\cite{sinha2013}. However, it is not known whether these are the highest possible magnetic fields. Ultra-strong fields of $10^{17}$--$10^{18}\Gs$ appear natural to consider in the context of compact stars, as suggested by simple dimensional analyses \cite{bratek2023} or inferred from a virial argument inside neutron stars \cite{lai1991}. Fields of $10^{19}\Gs$ may be present in the cores of compact stars \cite{yuan1998, tatsumi2000}. The upper limit for the magnetic field that  a quark star can sustain is of $10^{20}\Gs$~\cite{huang2010}, with a maximum field of $1.5{\times}10^{20}\Gs$ as follows from the upper limit for the magnetic energy density
imposed by the energy density of self-bound quark matter \cite{ferrer2010}.
The influence of strong magnetic fields is evident at the level of the equation of state for magnetized matter \cite{chakrabarty1996} and in the gravitational structure equations themselves. The magnetic field alters the form of the material stress-energy tensor and is included through the addition of the Maxwell fields' stress-energy part, thereby modifying the Oppenheimer-Volkoff-like equations.

In the previous work \cite{bratek2023}, the impact of ultra-strong magnetic fields on both neutron stars and strange stars composed of symmetric quark matter was examined. The study focused on the differences in responses of these two types of compact stars to such fields. It was confirmed that magnetic fields contribute to an increase in maximum masses for both neutron stars and strange stars. For neutron stars, the mass-radius relationship was qualitatively altered, whereas for strange stars, its shape remained similar to that observed in the absence of a magnetic field. The scaling relations characteristic of strange stars with symmetric matter were preserved in the particular model, even for fields as strong as $10^{18}\,\Gs$. In general, however, this result should depend on the model of coupling between the field and the star.

As in \cite{bratek2023}, the magnetic field is introduced into the structure equations at the level of the stress-energy tensor, with the simplifying assumption that the electromagnetic energy density is proportional to the energy density of SQM matter. This proportionality is motivated by the observation that in the case of symmetric SQM, the scaling symmetry with $\bag$ of the resulting TOV-like equations is preserved in the presence of a magnetic field \cite{bratek2023}. 
The approximation of exact spherical symmetry, utilized in this context \cite{bratek2023}, though inconsistent with globally divergent-free magnetic fields, allows for a simplified analysis of the gravitational effects of magnetic fields as encoded in the electromagnetic stress-energy tensor. While the local direction of realistic magnetic fields can vary greatly on smaller scales and be more ordered on larger scales, it is also assumed here that the directional character of the magnetic field is less crucial than its non-directional scalar characteristics, such as magnetic energy density, magnetic pressure, or the magnitude of magnetic tension.

The result of this work is the extension of scaling relations to SQM stars with magnetic fields, demonstrating how the symmetry is broken by the introduction of the s-quark mass. A phenomenological scaling formula for the masses of maximum mass stars is presented as a function of the surface magnetic field and s-quark mass. Additionally, an independent study of the mathematical structure of the SQM model with a massive s-quark is presented, developed concurrently with the numerical code tailored for this research. Baryon density is employed as the independent integration variable, a choice dictated by the numerical code's requirement to return the functions of state. This serves as an additional validation, illustrating that physically meaningful quantities describing a system such as a star remain consistent across different representations of the governing equations.

\section{Model assumptions and units conventions}

A set of dimensionless quantities along with their corresponding dimensional scales need to be introduced first, which 
will be used throughout. These are defined to simplify the equations so that they involve dimensionless quantities of 
order unity. This simplification ensures that the final equations are suitably prepared for further numerical computations.

\subsection{Dimensionless thermodynamical and stellar structure functions}

The number density of baryons will be described using a dimensionless function, $\nu$, which will serve as the independent variable in the equations of stellar structure, replacing the usual radial variable. This approach is feasible when $\nu$ is a monotonic function of the areal radius. In this way, the local pressure required to support the star against gravitational collapse will be directly determined from the current value of the integration variable $\nu$. The corresponding radial position in the star will be treated as a dependent function of $\nu$, to be found during the integration process.

  In the context of a spherical compact star described by material stress-energy tensor of perfect fluid,  the 
   following assumptions are made  for the thermodynamic mass density $\rho$, pressure $p$, and baryon 
   number density $n_B$,
   \begin{equation*}
   \begin{aligned}
   &\rho(r){=}\frac{p_o}{c^2}\mycal{R}(\nu(r/r_o)),\,\, p(r){=}p_o\mycal{P}(\nu(r/r_o)),\,\,\\
&n_B(r){=}n_o\nu(r/r_o),
\end{aligned}\end{equation*}
 with constant dimensional parameters $p_o$, $r_o$ and $n_o$ to be yet determined.
The choice of commensurate scales for $p$ and $\rho$ involves the linear combination $p+\rho c^2$ appearing 
in both thermodynamic and gravitational equations. 
{}
The resulting integrated mass and  particle number of the star,  on dimensional grounds, are expressed  as
$$ \mathcal{M}(r)=4\pi
\frac{p_o}{c^2} r_o^3\, \mycal{M}(\nu(r/r_o)), \quad \mathcal{N}(r)=4\pi
n_o r_o^3\, \mycal{N}(\nu(r/r_o)).$$

\noindent
For a particular fermion species contributing to $\rho$, $p$ and $\nu$, the scale of particle number density and the 
scale of mass density are chosen within the framework of an ideal fermion gas at a reference mass $m_r$  and 
its associated Compton wavelength. These scales are defined as follows:
\begin{equation}\label{eq:param}
\nref=\frac{1}{\pi^2}\br{\frac{\mref c}{\hbar}}^3,\qquad \epsref=\frac{3}{4}\mref c^2\nref.
\end{equation}
For a fermion regarded as massless, the reference mass 
 can be arbitrary, whereas for a fermion regarded as massive, it could simply be the fermion's mass. Additionally, 
 for each fermion species, a dimensionless particle density variable $\chi$ that incorporates a degeneracy factor 
 $\gamma$  will be used:
$$\chi_i(\nu)=\frac{6}{\gamma_i}\frac{n_i(n_B)}{\nref}=
\frac{6}{\gamma_i}\frac{n_o}{\nref}{\cdot}\nu{\cdot} f_i(\nu),\qquad f_i(\nu)=\frac{n_i(n_B)}{n_B}$$
where 
$\gamma=3$ for electrons and 
$\gamma=6$ for quarks.

\begin{table}[t]
\[
\begin{tabular}{@{}|@{\;}c@{\;}|@{\;}c@{\;}|@{\;}c@{\;}|@{\;}c@{\;}|}
\hline
$n_o$ & $\frac{1}{\pi^2}\br{\frac{m_f c}{\hbar}}^3$ & $8\pi{\cdot}\lambda_C^{-3}$ &
$0.04450560{\cdot}\br{\!\frac{m_f}{\mCL}\!}^{\!3}\,\fm^{-3}$ \\ 
$p_o$ & $\frac{3}{4}m_f c^2 n_o$ & 
$\frac{12\pi^2}{\alpha}{\cdot}e^2{\cdot} \lambda_C^{-4}$ &
$5.006880{\cdot}\br{\!\frac{m_f}{\mCL}\!}^{\!4}\,\frac{\MeV}{\fm^3}$ \\ 
$r_o$ & $\sqrt{\frac{c^4}{4\pi G p_o}}$ & 
$\frac{1}{\sqrt{48\pi^3}}\lambda_C^2/\mathcal{l}_P$ & 
$109.5708{\cdot}\br{\!\frac{m_f}{\mCL}\!}^{\!\!{-}2}\,\km$ \\ 
$m_o$ & $\frac{c^2r_o}{G}$ & 
$\frac{\mathcal{m}_P}{\sqrt{48\pi^3}}\lambda_C^2/\mathcal{l}_P^2$ & 
$74.18451{\cdot}\br{\!\frac{m_f}{\mCL}\!}^{\!\!{-}2}\,\Msun$ \\ 
$h_o$  & $\frac{q_o}{r_o^2}{=}\frac{c^2}{r_o\sqrt{G}}$ & $\sqrt{\frac{48\pi^3}{\alpha}}\frac{e}{\lambda_C^{2}}$ & 
$3.175000{\times}10^{17}{\cdot}\br{\!\frac{m_f}{\mCL}\!}^{\!2}\,\Gs$ \\
\hline
\end{tabular}
\]
\caption{\label{tab:units}Dimensional units characterizing a fermionic star, their scaling laws relative to the fermion mass $m_f$, and their values at the reference mass $\mCL=150\MeV/c^2$. These quantities are  expressed in terms of the Compton wavelength $\lambda_C=\frac{h}{m_f c}$, with 
$\mathcal{l}_P$ and $\mathcal{m}_P$ representing
the Planck length and Planck mass, $e$ the elementary charge, and $\alpha$ the fine structure constant.} 
\end{table}

The above form of mass function
and the general-relativistic scale $c^2/G$ for the mass-to-length ratio yield
the following scale parameters of length $r_o$ and mass $m_o$ at a specified scale of mass density  $p_o/c^2$. These parameters are derived from natural conditions within the framework of dimensional analysis for relativistic gravitating spheres,  involving only parameters $c$ and $G$:
$$4\pi p_or_o^2 = \frac{c^4}{G},\qquad {\rm and} \qquad \frac{Gm_o}{r_o} ={c^2}.$$ 
The first condition equates the integrated force of pressure $p_o$ over a sphere of radius $r_o$ to $c^4/G$, and the second links the Newtonian potential to $c^2$, with $m_o$ representing the mass of the sphere. 

\medskip\noindent
All units defined above and their scaling with the reference mass are summarized in \tabref{tab:units}.

 \subsection{Model of the stellar magnetic field contribution}

Electromagnetic fields contribute to the gravitational field equations through the electromagnetic component of the stress-energy tensor, introducing corrections to the energy density, along with anisotropic pressure and tension. This gravitational contribution can substantially affect the structure of a star, particularly by altering its total mass as perceived by remote test bodies.

In the context of electromagnetism in a strong gravitational field, a characteristic dimensional scale is provided by the charge-to-mass ratio which notably does not involve $c$:
$$\frac{q_o}{m_o}=\sqrt{G}.$$ 
But the nature of the electromagnetic charge this  implies is less clear. Stars cannot maintain strong, static, and spatially extended electric fields. This is because electric charge separation mechanisms capable of producing such fields are only feasible over very short distances within extremely thin spherical shells. On the other hand, stars may possess static and spatially extended strong magnetic fields that extend well beyond their surfaces, yet there are no magnetic charges to attribute to the charge-to-mass ratio. Moreover, the ratio $\sqrt{G}$
is vastly smaller than the anticipated ${\sim}10^{16}\sqrt{G}$ for a hypothetical magnetic monopole.
{This immense value is derived for the Dirac elementary
magnetic charge $\frac{e}{2\alpha}$ and a mass $\alpha^{-1}M_W$, where
$M_W{\sim}53\,\GeV{\cdot}c^{-2}$ is the upper limit for a typical vector
boson mass as considered in t’Hooft's field-theoretical
model of a magnetic monopole \cite{hooft1974}.} It is also further orders of magnitude smaller than for genuine electrically charged particles, such as ${\sim}10^{18}\sqrt{G}$ for the proton and ${\sim}2{\times}10^{21}\sqrt{G}$ for the electron. How, then, can one effectively make use of the virtually small $\sqrt{G}$? 

If it is so small, one might expect that $\sqrt{G}$ could define a physically viable scale for magnetic field strength\add{,} strong enough in the gravitational context, especially when considering the typical dimensional characteristics of compact stars, such as their size and mass. Interestingly, the associated scale, $h_o$, 
\begin{equation}\label{eq:hunit}
h_o=\frac{q_o}{r_o^2}
=\frac{m_o}{r_o^2}\sqrt{G}=\frac{1}{r_o}\frac{c^2}{\sqrt{G}}=\frac{1}{m_o}\frac{c^4}{G\sqrt{G}},
\end{equation}
while not as fundamental as $\sqrt{G}$ alone, has an order of magnitude comparable to that expected for compact stars with extremely strong magnetic fields: 
$$h_o
\approx \frac{10\,\km}{r_o}\cdot 3.479{\times}10^{18}\,\Gs=\frac{\Msun}{m_o}\cdot 2.355{\times} 10^{19}\,\Gs,$$ which gives $h_o=3.175{\times} 10^{17}\,\Gs$, see \tabref{tab:units}. Regardless of the adopted definitions for the parameters $m_o$ and $r_o$, which already include the constants $G$ and $c$, this scale can also be expressed in a complementary form that involves $c$ but not $G$:
$$h_o=\sqrt{\frac{m_o c^2}{r_o^3}}=\br{\frac{m_o}{\Msun}}^{1/2}\br{\frac{10\,\km}{r_o}}^{3/2}{\cdot} 1.337{\times}10^{18}\,\Gs.$$
Considering the aforementioned coincidence 
with the field values for compact stars, highlighted in the introduction, the subsequent question that arises is how to implement this scale in a simple magnetized compact star model within spherical symmetry. 

Although spherical symmetry is not compatible with divergent-free magnetic fields, it can be approximately applied to the scalars characterizing the electromagnetic stress-energy tensor, which includes magnetic strength and internal magnetization vectors. This tensor can be modeled similarly to that of a perfect fluid, but with the addition of locally anisotropic tension characteristic of electromagnetism, aligning compatibly with spherical symmetry. 

In this context, the magnetic-charge-to-mass ratio $\sqrt{G}$ can be considered; however, the resulting magnetic charges should be regarded as effective. The challenges of modeling magnetic fields under spherical symmetry and the issue of associated effective charges are discussed in more detail in \cite{bratek2023}. Following the model assumptions from \cite{bratek2023}, the effective magnetic charge density, $\rho_{mag}$, is associated with the material mass density, $\rho$, through the natural gravitational scale $\sqrt{G}$ of the mass-to-charge ratio. Specifically,  
\begin{equation}\label{eq:magncoupl}\rho_{mag}(r)=\kappa\sqrt{G}\rho(r), \quad
 \mathcal{Q}(r)=4\pi\frac{p_o}{c^2}r_o^3\sqrt{G}\,\mycal{Q}(\nu(r/r_o)),
\end{equation}
where $\mathcal{Q}$ represents the corresponding integrated effective charge.  A fractional dimensionless parameter, $\kappa$, is used to 
adjust the overall magnitude of magnetic field to the required value. 

\subsection{\label{sec:structeqs}Equations of the stellar structure and the electromagnetic mass contribution}

The equation of stellar equilibrium is traditionally expressed as a TOV-like equation for the radial gradient of pressure. However, parameterized with the particle density $\nu$, 
it can be reformulated as an equation for the radial variable 
{$ \mycal{S}{\equiv} \mycal{S}(\nu){:=}r(\nu)/r_o$}, 
while {$ \mycal{P}'{\equiv} \mycal{P}'(\nu)$} can be directly expressed by 
the equation of state. To implement this idea, an auxiliary function $ \mycal{W}$ and a shorthand {$\mycal{L}$} for the gravitational metric factor are introduced:  
\begin{equation*}
\begin{aligned}
 \mycal{W}&=
   -\frac{{  \br{ \mycal{M}
   + \mycal{P} \mycal{S}^3
   - \mycal{Q}^2/ \mycal{S}}\br{ \mycal{P}+ \mycal{R}}-\kappa { \mycal{L} \mycal{Q} \mycal{R}}}}{ \mycal{L}^2 \mycal{S}^2 \mycal{P}'}&
           \\  \mycal{L} &=\sqrt{1-\frac{2 \mycal{M}}{ \mycal{S}}+\frac{ \mycal{Q}^2}{ \mycal{S}^2}}.&
   \end{aligned}
   \end{equation*}
 In terms of these functions, the full stellar structure can be described by a system of first-order equations
\begin{equation}
\begin{aligned}\label{eq:structure}
&\mycal{Q}'=\kappa {\cdot}\frac{ \mycal{R}\, \mycal{S}^2}{ \mycal{L}\, \mycal{W}},\quad
\mycal{M}'=\br{1+\kappa\frac{ \mycal{Q}}{ \mycal{L} \mycal{S}}}\frac{ \mycal{R} \mycal{S}^2}{ \mycal{W}},\quad
\mycal{N}'=\frac{ \nu\, \mycal{S}^2}{ \mycal{L}\, \mycal{W}},\\
&\mycal{S}'=\frac{1}{ \mycal{W}},\quad
\mycal{U}'=\frac{ \mycal{M}+\mycal{P}\mycal{S}^3-\mycal{Q}^2/\mycal{S}}{\mycal{L}^2\mycal{S}^2\mycal{W}}.
 \end{aligned}
   \end{equation}
Here, the $'$ symbol denotes differentiation with respect to the particle density variable $\nu$, which is {regarded as} the independent variable. 
The thermodynamic functions $\mycal{R}$ and $\mycal{P}$ are directly expressed as functions of $\nu$ based on the particular form of the EOS. All remaining functions, including the radial variable, 
are{regarded as} functions of $\nu$ and will be determined by numerically solving this system.

 The above gravitational equilibrium equations 
 conform to the structure of the external Reissner–Nordström metric. This alignment allows the 'mass function' value at the star surface to overlap with the total stellar mass $M$, as ascertained by a remote test body moving in the star's gravitational field. Upon integration, these equations determine the total mass $M$ of the star as outlined in \cite{bratek2023}:
 \begin{equation}\label{eq:totalmass}
\begin{aligned}
&\mathcal{M}{=}\mathcal{M}_{\rm ext}{+}\int\limits_0^{R_{\star}}
 4\pi r^2\ud{r}\br{\rho(r){+}\frac{1}{c^2}\frac{\mathcal{H}^2(r)}{8\pi}}, \\ & \mathcal{M}_{\rm ext}=
\frac{R_{\star}^3\mathcal{H}^2(R_{\star})}{2c^2}{\approx}
\sq{\!\tfrac{\mathcal{H}(R_{\star})}{10^{18}\Gs}\!}^{\!2}\sq{\!\tfrac{R_{\star}}{10\km}\!}^{\!3}{\cdot}0.280\Msun. 
\end{aligned}\end{equation}
{Here, $\mathcal{H}$ is the magnetic scalar whose square determines the magnetic energy density.}
The above formula comprises two components:  the integral, accounting for both the material energy 
density and the magnetic scalar within the star, and a separate term representing the monopole field's contribution integrated from 
the stellar surface at $r{=}R_{\star}$ out to infinity (this formula might not be immediately clear because it involves integration over 
the entire space permeated by the magnetic field, whereas the integration of the structural equations is confined to the stellar interior only).
 
However, realistically, magnetic field exhibits a dipole structure outside the star.  Under a spherical approximation -- disregarding its angular dependence and assuming $\mathcal{H}(r){\approx} \mathcal{H}(R_{\star})(R_{\star}/r)^3$ for its radial decay -- the contribution of the external dipole field to the total mass would be approximately  $ \frac{1}{6c^2}R_{\star}^3\mathcal{H}^2(R_{\star})$,  effectively $1/3$ of the monopole contribution. The same value is obtained for the interior contribution, assuming a uniform scalar $\mathcal{H}{=}\mathcal{H}(R_{\star})$ for $r{<}R_{\star}$.   This crude estimation yields a total magnetic mass contribution of 
$$\mathcal{M}_{\rm mag}\approx\frac{R_{\star}^3\mathcal{H}^2(R_{\star})}{3c^2}{\approx}\sq{\!\tfrac{\mathcal{H}(R_{\star})}{10^{18}\Gs}\!}^{\!2}\sq{\!\tfrac{R_{\star}}{10\km}\!}^{\!3}{\cdot}0.186\Msun.$$

\section{Revisiting the mathematical structure of the SQM model with a massive s-quark}

The pivotal concept of the SQM model is the bag constant $\bag$. 
A value of $58\, \MeV{\cdot}\fm^{-3}$ denoted by $\bag_{58}$ is frequently used  as a
 reference value for $\bag$. Here, another value is adopted:   $$\bagref=57.81068\, \MeV{\cdot}\fm^{-3}. $$ 
This value is chosen because then the symmetric SQM star with $\bag=\bagref$ has a maximum mass of $2\Msun$ 
(assuming $\Msun=1.98892\times10^{33}{\gram}$). 
Magnetic field with the energy density equal to $\bagref$ would correspond to the field strength:
\begin{equation}\label{eq:magbagunit}
\magref\equiv \sqrt{8\pi\bagref}=1.525735\times 10^{18}\Gs, 
\end{equation}
 which provides a convenient unit.

The thermodynamic state of SQM matter with a massive s-quark is determined by the baryon number density $n_B$ at a given value of the bag constant $\bag$ and the s-quark mass $m_s$, which are the dimensional parameters of the model. The value of $n_B$ affects the relative number density of quarks  $u=n_u/n_B$, $d=n_d/n_B$, $s=n_s/n_B$ and that of electrons $e=n_e/n_B$ which are functions of $n_B$ (they would remain constant for massless s-quark, when SQM becomes symmetric: $u{=}d{=}s{=}1$ and  $e{=}0$). 
In the units adopted here, the baryon number density will be described in terms of its dimensionless analogue $\nu\equiv n_B/\nref$, where $\nref$ is the number density scale defined at a universal reference mass $m_r$. The $m_r$  will be used to define also other dimensional scales, such as the energy density or pressure, as discussed earlier. In the present context, $m_r$ will be equated with the s-quark mass $m_s$, while all other particles will be regarded as massless. This way, the model becomes characterized by only a single 
dimensionless numerical parameter $b$ which expresses the bag constant $\bag$ in the adopted units of energy density defined at the assumed s-quark mass $m_s$, namely
\begin{equation}\label{eq:bdef}b= b(\bag,m_s)\equiv\frac{\bag}{\epsref}=
11.98351{\cdot}\frac{\bag}{\bag_{60}}{\cdot}\br{\frac{m_{150}}{m_s}}^4 
\end{equation} with $\bag_{60}{\equiv}60\,\MeV{\cdot}\fm^{-3}$,
$m_{150}{\equiv}150\,
\mathrm{MeV}{\cdot}c^{-2}$.
All other changes in the model go through simple rescaling of units.

\subsection{Degenerate fermion gas}

The theoretical basis of the SQM model with a massive s-quark is simple, as it is furnished by the formulas for degenerate fermion gas. Nevertheless, using the model in practice is numerically more demanding. 

For a fermion gas, 
the thermodynamical functions, such as the the particle number density $n$, the energy density $\epsilon$, the pressure $p$, and the chemical potential 
$\mu$ become functions of each other. They will be expressed through the number density parameter $\chi$ at a mass scale $m_r$, as defined in \tabref{tab:degenerate}. 
\begin{table}[!th]
\[ \begin{tabular}{|@{\;}c@{\;}|c|c|c|c|}
\hline
 & $n/\nref $ & $\epsilon/\epsref$ & $p/\epsref$ & $\mu/\mref c^2$ \\
\hline
\raisebox{-1.2ex}{{\small\shortstack{massless\\fermion}}}  & $ \frac{\gamma}{6}\chi$ 
& $ \frac{\gamma}{6}\chi^{4/3}$ & $\frac{1}{3}{\cdot}\frac{\gamma}{6}\chi^{4/3}$ & $\chi^{1/3}$\\
\hline
\raisebox{-1.2ex}{{\small \shortstack{massive\\ fermion}}}  & $ \frac{\gamma}{6}\chi$ & $\frac{\gamma}{6}K(\chi^{1/3})$ 
& $\frac{1}{3}{\cdot}\frac{\gamma}{6}J(\chi^{1/3})$ & $\sqrt{1+\chi^{2/3}}$\\
\hline
\end{tabular}
\]
\caption{\label{tab:degenerate}Principal quantities characterizing a degenerate gas of noninteracting fermions in units defined at the same reference mass scale $m_r$ equal to the mass of the massive fermion.  The degeneracy factor $\gamma= 2{\cdot}3$ for quarks and $\gamma=2$ for electrons. }
\end{table}
Functions $K$ and $J$ appearing in the table are defined as 
\relax
\begin{equation}\label{eq:KJdef}
\begin{aligned}
&\!\!\!\! K(\xi)\equiv\frac{1}{2}\sq{\br{1+2\xi^2}\xi\sqrt{1+\xi^2}- \ln{\br{\xi+\sqrt{1+\xi^2}}}}, \\
&\!\!\!\! J(\xi)=\frac{1}{2}\sq{\br{2\xi^2-3}\xi\sqrt{1+\xi^2}+3\ln{\br{\xi+\sqrt{1+\xi^2}}}},
\end{aligned}
\end{equation}
where, $J(\xi)\equiv\xi K'(\xi)-3K(\xi)$. 
\relax
The reference mass $m_r$ has been equated with the massive fermion mass both for massive and massless fermion case. With this choice of units, 
   the thermodynamical quantities  in the same column of table \ref{tab:degenerate} can be compared with each other, in particular, they can be seen to converge one to another in the large particle density limit, as can be seen from the  
asymptotic expansions at $\chi=\infty$:
\begin{equation*}
\begin{aligned}
{\chi}^{-4/3}K({\chi }^{1/3})&=1+\chi^{-2/3}\\
-&(1/3) \left( \ln{\sqrt{8\chi}}-3/8\right)\chi^{-4/3}+\mathcal{O}(\chi^{-2}),\\
{\chi}^{-4/3}J({\chi }^{1/3})&=1-\chi^{-2/3}\\ 
+&\left( \ln{\sqrt{8\chi}}-7/8\right)\chi^{-4/3}+\mathcal{O}(\chi^{-2}).
\end{aligned}
   \end{equation*}
   
\subsection{SQM at the minimum of microsopic energy}

The material energy density $\mathcal{R}$ in the considered SQM model consists of separate contributions from a non-interacting mixture of degenerate gases: massive s-quarks,  massless u-quarks and d-quarks, and massless electrons, with an additional effective contribution introduced by a bag constant $\bag$:
$$\mathcal{R}(n_u,u_d,n_s,n_e)=\epsilon(n_u)+\epsilon(n_d)+\epsilon(n_s)+\epsilon(n_e)+\bag.$$ The form of 
$\epsilon(n_i)$ for each particle species 
is detailed in \tabref{tab:degenerate}. Specifically,  $\epsilon(n_s)=\epsref{\cdot}K((n_s/\nref)^{1/3})$ for massive s-quarks, $\epsilon(n_i)=\epsref{\cdot}(n_i/\nref)^{4/3}$ for massless d-quarks and u-quarks, and $\epsilon(n_e/n_r)=3^{1/3}\epsref{\cdot}(n_i/\nref)^{4/3}$ for massless electrons. It is important to emphasize that the reference mass $m_r$ has been set equal to the s-quark mass $m_s$, that is $m_r=m_s$ for all particle species entering the above formula for $\mathcal{R}$.

Furthermore, in every state, SQM should be considered at its minimum energy in a given volume element.  
The condition of minimum energy is constrained by the requirement of charge neutrality $+(2/3)n_u-(1/3)n_d-(1/3)n_s-n_e=0$  at a given value of 
baryon number density
$n_B-(1/3)n_u-(1/3)n_d-(1/3)n_s=0$. These two conditions act as primary constraints. Any  equivalent combination
of these constraints can also be used -- here, the combination $n_u-n_B-n_e=0$ 
replaces the charge neutrality constraint, leading to a conditional minimum problem:
\begin{equation*}
\begin{aligned}
f=&\mathcal{R}(n_u,u_d,n_s,n_e)\\+&\lambda_B(3n_B-n_u-n_d-n_s)
+\lambda_Q(n_u-n_e-n_B).
\end{aligned}
\end{equation*}
 This yields 
a set of $6$ equations for $6$ variables: $n_u,n_d,n_s,n_e,\lambda_B,\lambda_Q$. These consist of the pair of primary constraints and four stationary point conditions: $\mu_u-\lambda_B+\lambda_Q
=0$, $\mu_d-\lambda_B=0$, $\mu_s-\lambda_B=0$, $\mu_e-\lambda_Q=0,$ where the definition of chemical 
potentials $\mu_i\equiv\epsilon'(n_i)$ has been used.
However, the rank of the set of equations is deficient, with a rank of $4$ instead of $6$. This deficiency introduces two additional (secondary) constraints, resulting in the  following conditions for the chemical potentials
$$\mu_s=\mu_d=\mu_e+\mu_u.$$
 Then, the remaining solutions are $\lambda_Q=\mu_e$, $\lambda_B=\mu_s$ (of no further interest) and conjugated 
 to them the original pair of primary constraints.  
 Now, it is left to set $\gamma_i=2{\cdot}3$ for quarks and $\gamma_e=2$ for the electron,  and to introduce the relative densities as defined earlier, while assuming the same reference mass scale $m_r=m_s$ for all particle species. Finally,  
the expressions for $\mu_i$ as functions of $n_i$, along with the four constraints, give:
\begin{equation}\label{eq:surface}
\begin{aligned}
&u-e=1,\quad u+d+s=3,\quad d^{1/3}=(3e)^{1/3}+u^{1/3}\\
&\quad 1+(\nu s)^{2/3}=
 (\nu d)^{2/3},
\end{aligned}\end{equation}
 
\subsubsection{\label{sec:qualdis}Qualitative discussion of the minimum energy solution}
  
Given a $\nu$ value, the relative densities $u,d,s$, which must be positive, attain the corresponding unique values  $s<1<u<d<2$ for $e>0$. However, $\nu$ cannot be arbitrary and is limited from below by a positive number, $\nu_m$. Otherwise, the $3$rd equation in \eqref{eq:surface} would imply that either $d$ or $s$, or both, would diverge as $\nu\to0$, which contradicts the $2$nd equation. Furthermore, taking the opposite limit, as $\nu\to\infty$, the $4$th equation implies that $s\to d$, resulting in $u{=}d{=}s{=}1$ and $e=0$, as follows from the remaining equations in \eqref{eq:surface}. This reinforces the expectation that for sufficiently large $\nu$, the prediction of the model with massive s-quark should align with the symmetric SQM model. In the latter case, obtained for the massless s-quark case, there would be $0$ instead of $1$ in the $4$th equation in \eqref{eq:surface},resulting in $e=0$ and  $u{=}d{=}s{=}1$, with unconstrained $\nu$ acting as an independent density parameter at an arbitrary mass scale $m_r$.
Then $\mathcal{R}=3\nu^{4/3}+b$ at this scale. 
Assuming a massless s-quark from the start, a similar calculation leads to the same conclusion.
 
 \subsubsection{A parametric form of the minimum energy solution}

Having established the principal equations \eqref{eq:surface}, one needs to determine 
the resulting fractional densities $u,d,s,e$ as functions of the baryon density $\nu$. Unfortunately, 
finding an explicit form of these variables is not feasible. Instead, a parametric description
will be utilized, 
with one of the variables defined as a function of an auxiliary parameter $x$, and the remaining functions expressed accordingly, 
so that \eqref{eq:surface} is identically satisfied for any $x$. 
 In particular, the fractional electron density $e$ serves as a useful parametrization. 
 The choice to set  $e(x){=}x^3$ here,  is driven by the observation that the Taylor series for the function  $x^{3/2}\nu(x)$ exclusively comprises non-negative integer powers of $x$.
With this parametrization, the parametric solution of \eqref{eq:surface}
takes the following form:
{}
\begin{equation}\label{eq:parametric}
\!\!\!\!\!\!\!\!\left.
\begin{array}{l}
e(x)  = x^3\\
u(x)  = 1{+}e(x) =1{+}x^3\\
d(x)  = \br{\!\sqrt[3]{3}e^{1/3}(x){+}u^{1/3}(x)\!}^{\!3}=(\sqrt[3]{3}\,x{+} \sqrt[3]{1+x^3})^{3}\\
s(x)  = 2{ -}e(x) {-} d(x) =2{ -}x^3 {-} (\sqrt[3]{3}\,x{+} \sqrt[3]{1+x^3})^{3}\\
\end{array}
\right.
\end{equation}
and now the last expression in  \eqref{eq:surface} can be solved for the density $\nu$, which gives
\begin{equation*}
\begin{aligned}
&\nu(x) = \big(d^{2/3}(x)-s^{2/3}(x)\big)^{-3/2}
=\\&\frac{1}{\big(
\big[\!\sqrt[3]{3}x{+}\sqrt[3]{1+x^3}\,\big]^2{-}
\big(
{
2 {-}x^3 {-}
\big[
\sqrt[3]{3}x{+}\sqrt[3]{1+x^3}
\,\big]}^3
\big)^{2/3}
\big)^{3/2}}.
\end{aligned}
\end{equation*}
The resulting Taylor series
 $$8\,{\sqrt{3}}\,x^{3/2}\,\nu (x) = 
  1 - \frac{3}{2}\,3^{\frac{1}{3}}\,x + 
   \frac{1}{8}\,3^{\frac{2}{3}}\,x^2 + \ldots, \quad 0<x<x_m$$ 
 yields  to the corresponding inverse series, suitable at large $\nu$, $\nu\gg\nu_m$ 
  (where $\nu_m\equiv \nu(x_m)$ and  $x_m$ is a limiting value corresponding to $\nu_m$ already alluded to 
  earlier in \secref{sec:qualdis}), namely, 
   \begin{equation}\label{eq:inverseseries1}
   \begin{aligned}
   &x(\nu)=
   &\frac{1}{4\sqrt[3]{3}\nu^{2/3}}
   \br{1- \frac{1}{4}\frac{1}{\nu^{2/3}}+\frac{5}{96}\frac{1}{\nu^{4/3}}+\ldots}.\end{aligned}\end{equation}
    As $x$ increases from $0$, $u(x)$ and $d(x)$ both rise, hence $s(x)$ decreases,  ultimately reaching $0$ at some 
    $x=x_m$. Equation $s(x_m)=0$ is solved by
     $$x_m^3{=}{e_m{=}}(9t_m^{1/3}{-}72t_m^{-1/3}{-}1)/22,\quad t_m{\equiv}(3{+}11\sqrt{17})/2,$$ which gives 
     $x_m{=}0.17809294\cdots$ (several constants corresponding to this limiting value are shown in \tabref{tab:param}).
\begin{table}[ht]
\[   \begin{tabular}{r|r}
   \hline
    $b_m=\nu_m^{4/3}\Pi(\nu_m)$ & $0.4673102$\\
    \hline
   $x_m$ & $0.1780929$\\
   $e_m$ & $0.005648591$\\
  {$d_m$} & {$1.994351$}\\
   $\nu_m=\nu(x_m)$ & $0.5014161$\\
    $\Pi_m=\Pi(\nu_m),\quad  \Epsilon_m= \Epsilon(\nu_m)$ & $ 1.173116$\\
      $d_m/u_m$ & $1.983149$\\
   $\gamma_m=- (2/3)^{3/2} d_m/s'(x_m)$ & $0.1528847$\\
   \hline
   $R_m/\nu_m$ & $3.727923$\\
   \hline
   \end{tabular}
\]   
 \caption{\label{tab:param}
 Critical values of selected quantities in the SQM model with a massive s-quark {at $s{=}s_m{\equiv}0$}. 
{They} are independent of both the bag constant and {the} s-quark mass {free parameters}.\medskip\\
{ The defining 
 expressions for the constants in the left column of \tabref{tab:param} can be reduced  and then expressed in exact form 
 by setting $s{=}0$ prior to substituting $x{=}x_m$. 
 In particular, $b_m{=}\frac{2}{81}(13{+}\tau_m^{1/3}{+}4\tau_m^{-1/3})$ with $\tau_m{=}(137{-}33\sqrt{17})/2$.}
}
\end{table}

At $x=x_m$, matter {decreases to a limitng} density  at which the s-quarks disappear 
and  below which SQM cannot exist. This threshold value 
{corresponds to a lower bound for the  baryon density} determined solely by the s-quark mass 
   $$n_m(m_s)= \nu_m \nref =0.02231582\, \fm^{-3}\cdot  \br{\!\frac{m_s}{\mCL}\!}^3.$$ At this density,  
the ratio $d(x)/u(x)$ attains {its maximum} value of $d_m/u_m{\approx}1.983${, very} close to 
that for isolated neutrons, which is $2$.   

\subsubsection{\label{sec:pressure}Pressure as the Legendre transformation of the energy density. Connection with partial pressures.}

In the idealized situation described earlier, there is a mixture of several particle species in a steady state of minimum energy, and these species are assumed not to interact directly with each other (apart from the interactions encapsulated in the imposed conservation constraints). Thus, one expects the subsystems to contribute additively to the total energy density and pressure. The composite pressure $\mathcal{P}$ could be thus regarded as the mere sum of the individual (partial) pressures determined independently for each particle species, reduced by the bag constant: $$\mathcal{P}=p(n_u)+p(n_d)+p(n_s)+p(n_e)-\bag.$$
These individual pressures are the Legendre transformations of the individual energy densities with respect to the corresponding particle densities:  $p_i{=}{-}\epsilon_i{+}n_i{\cdot}\partial_{n_i}\!\epsilon_i$. On the other hand, the overall pressure for this mixture of subsystems could be defined as the Legendre transformation of the total energy in an infinitesimally small volume element (expressed in terms of local energy density $\mathcal{R}$) with respect to the total baryon number in that volume (expressed in terms of local baryon number density $n_B$), $$\mathcal{P}{=}{-}\mathcal{R}{+}n_B{\cdot}\partial_{n_B}\mathcal{R}.$$ 

\noindent
It turns out that these two methods of computing pressure indeed yield identical results when the energy per unit volume attains its minimum at a given baryon number density. This can be demonstrated as follows:
By calculating the derivative of the energy density, restricted to be a function of fractional densities only, one obtains an expression that evaluates identically to $0$: 
\begin{equation*}
\begin{aligned}
&\frac{\nu}{3}\left[(\nu s)^{1/3}K'((\nu s)^{1/3})\frac{s'(\nu)}{s}+4(\nu u)^{4/3}\frac{u'(\nu)}{u}\right.\\
&\left.+4(\nu d)^{4/3}\frac{d'(\nu)}{d}+4{\cdot} 3^{1/3}(\nu e)^{4/3}\frac{e'(\nu)}{e}\right]=0,
\end{aligned}
\end{equation*}
as follows from the formulas in \tabref{tab:degenerate} and the previous results. This is the condition required to show that the two methods of calculating the pressure are equivalent. Indeed,
$\mathcal{P}=\nu{\cdot}\mathcal{R}'(\nu)-\mathcal{R}=\sum_i[\nu{\cdot}n_i'(\nu){\cdot}\epsilon'_i(n_i)-\epsilon_i]=
\sum_i[(\nu^2{\cdot}s_i'(\nu)+\nu{\cdot}s_i){\cdot}\epsilon'_i(n_i)-\epsilon_i]=
\sum_i[n_i{\cdot}\epsilon'_i(n_i)-\epsilon_i]+\nu^2{\cdot}\sum_i\epsilon'_i(n_i) {\cdot} s_i'(\nu)
=\sum_i p_i+\nu{\cdot}\sum_i\partial_{s_i}\epsilon_i(s_i\nu){\cdot} s_i'(\nu)$. Hence, for the equality $\mathcal{P}=\sum_i p_i$ 
to hold one needs 
$\sum_i\mu(n_i){\cdot} s_i'(\nu)=0$, taking that $\partial_{s_i}\epsilon_i(s_i\nu)=\nu\mu(n_i)$, 
 at a given $\nu$ at the minimum energy density. This implies that the energy per given volume element does not change when particles are transformed within that element, as long as the $\nu$ remains unchanged. The direction 
 of the differentials vector $\ud{s}_i$ is not arbitrary, but precisely such that the flow of energy between particle species 
 does not affect the total energy in that volume element -- it is the same as otherwise resulting from the change of 
 $\nu$ in a transition to another minimum energy density state at a slightly different $\nu$.
This is intuitively clear -- at its minimum the energy density is not altered when fractional densities are varied in such 
a way that the particle density $\nu$ is kept constant. 

In a more complex scenario, the neglected interactions between subsystems could affect the energy density and thus the pressure of the mixture. The overall pressure in this case would not simply be the sum of the individual pressures, as the interactions could contribute additional terms to the total pressure of the system.

\subsection{\label{eq:eoslim}The equation of state of SQM with massive s-quark}

\begin{figure}[htb]
\centering
\includegraphics[trim=0 0 0 0,clip,width=0.8\columnwidth]{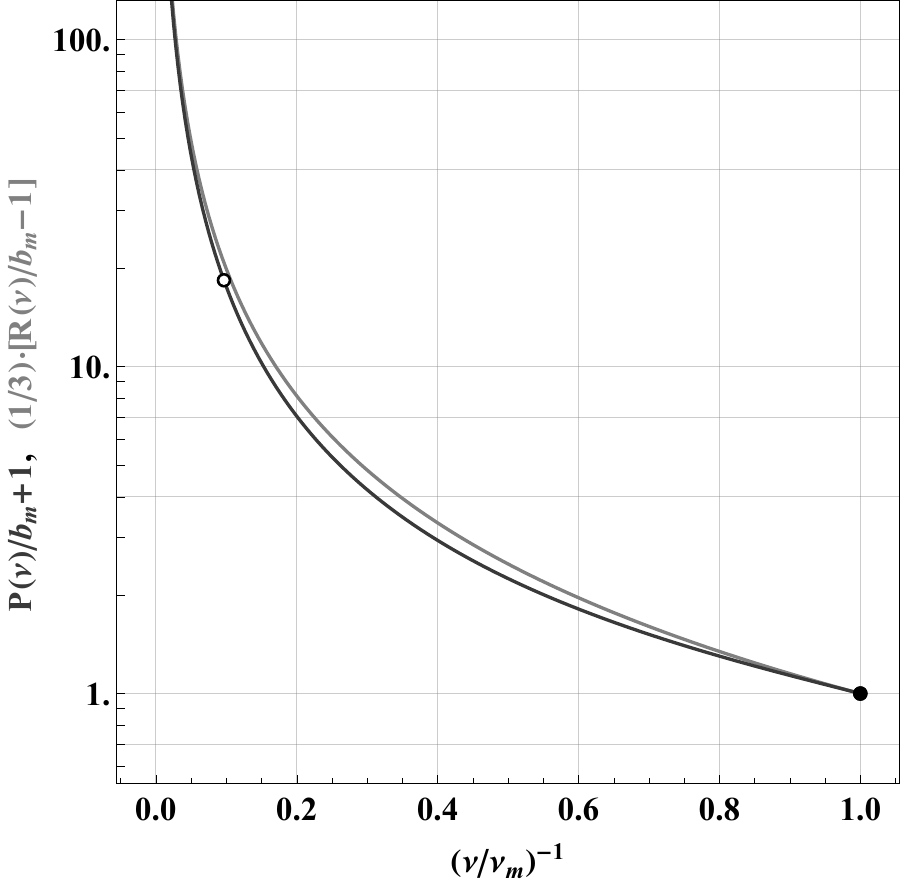}\\
\includegraphics[trim=0 0 0 0,clip,width=0.8\columnwidth]{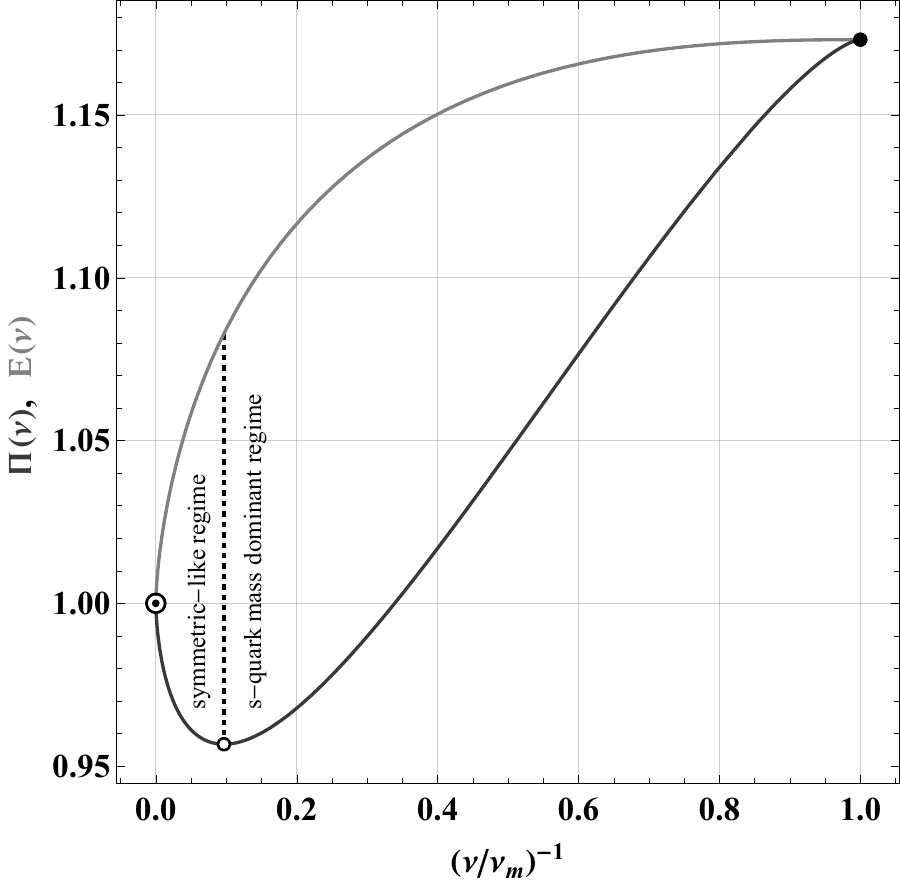}
\caption{\label{fig:criticalb}
The behavior of the pressure and energy density functions for the SQM equation of state with a massive s-quark, 
$\Pi$ and $P$ (dark gray lines), and $\Epsilon$ and $R$ (light gray lines),  
are shown against the standardized proper volume per baryon $(\nu/\nu_m)^{-1}$. In both plots, 
the {small} open black circle indicates the position of the minimum of $\Pi$. 
The filled black circle corresponds to the value at the threshold density $\nu=\nu_m$, below which s-quarks cease to exist. 
The {center--}\unskip dotted circle 
represents the values at the theoretical limit of infinite density {(the symmetric SQM case)}.
 }
\end{figure}

In numerical computations of the equation of state, it is beneficial to introduce a pair of bounded auxiliary functions defined as follows:
\begin{equation}\label{eq:EPidef}
\begin{aligned}
\Epsilon(\nu)&{=}\br{1/3}{\cdot}
\left[
\nu^{-4/3}{\cdot}K([\nu{\cdot}s(\nu)]^{1/3}) 
\right.\\
&
\left.+[u(\nu)]^{4/3}+[d(\nu)]^{4/3}+3^{1/3}{\cdot}[e(\nu)]^{4/3} \right]\\
\Pi(\nu)&\equiv\Epsilon(\nu)+3\nu\Epsilon'(\nu)=
\br{1/3}{\cdot}
\left[
\nu^{-4/3}{\cdot}J([\nu{\cdot}s(\nu)]^{1/3})\right.\\
&\left.+[u(\nu)]^{4/3}+[d(\nu)]^{4/3}+3^{1/3}{\cdot}[e(\nu)]^{4/3} \right],
\end{aligned}
\end{equation}
where $K$ and $J$ were defined in \eqref{eq:KJdef}. 
Using the functions $\Pi(\nu)$ and $\Epsilon(\nu)$, 
the total energy density and pressure can be expressed in a dimensionless form: 
\begin{equation}\label{eq:zwiazkifunkcji}
\left.\begin{array}{l}
R(\nu)  \equiv \mathcal{R}(\nref \nu)/\epsref=3\,\nu^{4/3}\Epsilon(\nu)+b,\medskip\\
P(\nu)  \equiv \mathcal{P}(\nref \nu))/\epsref=\nu R'(\nu)-R(\nu)=\nu^{4/3}\Pi(\nu)-b. 
\end{array}\right.
\end{equation}
It is understood that all functions of $\nu$ in these equations are computed indirectly through the inverse function to $\nu(x)$, using the parametric description \eqref{eq:parametric}. 
The behavior of these functions is illustrated in \figref{fig:criticalb}. 
For comparison, in the massless s-quark case where $e=0$ and $u=d=s=1$ for any $\nu$, the functions simplify to $\Pi=\Epsilon=1$, $K=J=\nu^{4/3}$.
Subsequently, the complete equation of state can be put in a form that isolates the component characteristic of 
the massless equation of state:
\begin{multline}
\label{eq:eosmain}
R(\nu)-\sq{3P(\nu)+4b}=K-J=3\nu^{4/3}\br{\Epsilon(\nu)-\Pi(\nu)}=\\
2\nu^{2/3}\bigg(1-\frac{ \ln \left({8 \nu}\right)}{3\nu^{2/3}} +O\big({\nu^{-4/3}
   }\big)\bigg).\end{multline}

It can be verified that the right-hand side of the expression $R-3P-4b=K-J$ in  \eqref{eq:eosmain}
 is an increasing 
function of its argument $y=(s\nu)^{1/3}$ (cf. \eqref{eq:EPidef}), starting from a value of $0$ at $y=0$ and 
exhibiting leading behaviors of  $\sim\frac{4}{3}y^3$ as  $y\to0$ and
  $\sim 2y^2$ as $y\to\infty$. Thus, $R> 3P+4b$ holds true  whenever $s>0$.

   The asymptotic behavior of $K-J$ for large $y$ means that $R\approx3P+4b$ at high s-quark densities,
 and hence to high baryon densities. Meanwhile, the contribution from all particles species to 
  $R$ not accounted for in the $K-J$ part behave as $\nu^{4/3}\sim y^4$ and dominate the $y^2$ behaviour.
  Thus, at sufficiently high baryon densities, the model resembles 
  the symmetric model with the equation of state $R=3P+4b$. For moderate $y$, the energy density 
  is increased compared to the symmetric model,  $R>3P+4b$. Finally, as $y\to0$, again one finds 
  $R\approx3P+4b$, with equality reached at $y=0$  (the analytical result $K=J$ at $y=0$
   explains the numerical coincidence $\Pi_m=\Epsilon_m$ displayed in 
  \tabref{tab:param}, where
    $\Pi_m\equiv\Pi(\nu_m)$ and $\Epsilon_m=\Epsilon(\nu_m)$).    
    
      Since $\nu\geqslant\nu_m>0$, the equality can occur only when $s=s_m=0$, indicating that 
      $\nu$ has reached $\nu_m$. 
   This is only possible
if $b$, and consequently the ratio $\bag/m_s^4$, is reduced to align with the critical ratio $b_m$.
When $b=b_m$,
the s-quark mass reaches its maximum possible value, while the s-quark fraction $s$ vanishes 
(meaning the s-quarks disappear), and the electrons' fractional contribution $e$ reaches its maximum value $e=e_m$.

This behavior suggests that the contribution from electrons is essential for SQM with a massive s-quark. However, 
their correction to the energy density of other massless 
fermions can be neglected for small s-quark masses. In this case, the model is encompassed 
 within a simpler unifying framework considered by  \citet{zhang2021}
 to first order in 
$m_s^2$, neglecting the electron contribution to the energy density. 
To estimate the electron contribution in the present model in the limit of small $m_s$, 
one can introduce $\mu$ as the expansion parameter 
with respect to $m_s$, and perform an expansion in
$e$ and $\mu$, treating the product $\mu{\cdot}\nu(e)$ as constant, and disregarding terms of order $o(e^{2/3})$ and $o(\mu^2)$.
This leads to
$$R{\approx}(3{+}4(3e)^{2/3})\nu^{4/3}+\mu^2(1{-}2(3e)^{1/3}{-}3(3e)^{2/3})\nu^{2/3},$$
where $\mu$ must be set to $1$ in the adopted units.
In this order of expansion in $\mu$, the correction from the electron fractional density $e$ to the symmetric baryonic 
component $3\nu^{4/3}$ and
to the s-quark mass term component $\mu^2\nu^{2/3}$ is indeed negligible, since $e<e_m\approx0.0056$ as proved earlier. 
However, this is not the case in the extreme scenario when $s\to0$, for which the low order expansion is invalid.

The condition
$b=b_m$ also indicates that the saturation density has reached its lower bound.     
   At the saturation point, $0=(R/\nu)'$, which implies $0=P/\nu^2$ from 
the defining property $P\equiv\nu R'-R$. Hence $P=0$ at that point, meaning that SQM saturates  
at the star's surface. Let all quantities at that point be denoted with the subscript $\sigma$.
The condition $P\sat=0$ leads to a highly nonlinear equation for 
$\nu=\nu\sat(b)$:
$\Pi\sat\nu^{4/3}\sat=b$. This equation becomes singular near $\nu_m$ and must be solved at the quark star surface. 
Then, $R\sat=(\Pi\sat+3\Epsilon\sat)\nu^{4/3}\sat>4 b$ 
if $s_{\sigma}>0$, 
as derived from \eqref{eq:zwiazkifunkcji}.
 Conversely, assuming the limiting case $s_{\sigma}=0$, one finds $\nu_{\sigma}=\nu_m$, which is only possible if 
  $b=b_m$. In this case, $R_m=4b_m$ from the definition of $b_m$ and the property $\Pi_m=\Epsilon_m$ shown earlier. 
     
In the limit of a massless s-quark, the adopted scales of pressure, energy density, and particle density all 
shrink to zero. Therefore, {before} taking this limit, both sides of 
{\eqref{eq:eosmain}} must be multiplied by 
$(m_s/m_{150})^4$ and re-expressed in terms of the rescaled quantities 
($\tilde{P}= (m_s/m_{150})^4P$, $\tilde{R}= (m_s/m_{150})^4R$ and 
$\tilde{\nu}\equiv\nu{\cdot} (m_s/m_{150})^3$)\add{,} with units defined at the $m_{150}$ scale. 
In this scenario, as $\Epsilon$ and $\Pi$ both {approach} $\Pi(\infty)=\Epsilon(\infty)=1$ in the limit 
{as} $m_s\to0$, 
the right-hand side of the equation tends to zero, \repl{thus}{thereby} re-establishing the equation of state for SQM 
with a massless 
s-quark\add{:} $$\tilde{R}(\tilde{\nu})=3\tilde{P}(\tilde{\nu})+4\tilde{b},$$ as expressed in units at the $m_{150}$ scale.

\subsubsection{\label{sec:criticalbag}Critical values of the bag constant  and the s-quark mass in the SQM model}
In the stellar interior, the pressure $P$ remains positive but reaches zero at the star's surface. At this saturation point, 
the energy per particle, $R/\nu$, attains a local minimum. This follows from the identity  
$\nu^2\partial_{\nu}(R/\nu)=\nu\partial_{\nu}R-R\equiv P$, and from the fact that for SQM, $\nu>0$ even when $P=0$. 
Consequently,  $\partial_{\nu}(R/\nu)|_{\rm sat}=0$, and furthermore,  
$\partial^2_{\nu}(R/\nu)|_{\rm sat}=\nu^{-2}\partial_{\nu}P|_{\rm sat}=\nu^{-3}\beta^2R|_{\rm sat}>0$, 
as required 
for a local minimum (here, $\beta^2=\partial_{\nu}P/\partial_{\nu}R$ is the isentropic sound velocity 
squared { expressed in units of the speed of light}). 

For a massless s-quark, the condition $P>0$ can be met with any $\bag>0$, provided 
the baryon density exceeds the saturation value 
{$\nu|_{\rm sat}{=}b^{3/4}$, or in the dimensional form $0.2866505\,\fm^{-3}{\cdot}\br{\frac{\bag}{\bag_{60}}}^{3/4}$}, 
which corresponds to the minimum energy per baryon, 
{$R/\nu|_{\rm sat}{=}4b^{1/4}$ or $837.2565\,\MeV{\cdot}\br{\frac{\bag}{\bag_{60}}}^{1/4}$}.  
However, in the SQM model with a massive s-quark, 
the second equation in \eqref{eq:zwiazkifunkcji} identifies a critical value of $b$ below which the condition  $P>0$ cannot be satisfied  (see \figref{fig:criticalb}). 
This is because $\nu^{4/3}\Pi(\nu)$ is a non-decreasing function of $\nu$, as 
$\partial_{\nu}(\nu^{4/3}\Pi(\nu))=\partial_{\nu}P\equiv \beta^2\partial_{\nu}R=\beta^2\frac{P+R}{\nu}>0$. 
Particularly\add{,} at the saturation density $\nu=\nu|_{\rm sat}>\nu_m$ when $P=0$, this property results in $b=\nu^{4/3}\Pi(\nu)|_{\rm sat}>\nu_m^{4/3}\Pi(\nu_m)\equiv b_m$, 
establishing $b_m$ as a  critical {lower bound} value (see \tabref{tab:param}). 
Thus, the equation $b=\nu^{4/3}\Pi(\nu)|_{\rm sat}$ can only be solved if $b>b_m$; specifically, if
 $$\frac{b}{b_m}=25.64359 {\cdot}\frac{\bag}{\bag_{60}}{\cdot}\br{\frac{m_{150}}{m_s}}^4{>}1, \quad b_m\equiv \nu_m^{4/3}\Pi(\nu_m),$$  or in terms of the critical bag constant $\bag_m$,  
  \begin{equation*}\begin{aligned}\bag>\bag_m&\equiv 2.339766\, \MeV{\cdot}
\fm^{-3} \br{\frac{m_s}{\mCL}}^4,\\ &{\rm or}\quad m_s<337.5483\, \MeV/c^2{\cdot}\br{\frac{\bag}{\bag_{60}}}^{1/4}.\end{aligned}\end{equation*}
At $\bag=\bag_m$, the saturation energy per particle $R/\nu|_{\rm sat}$ attains its lowest possible value in this model of $419.3913\,\MeV{\cdot} \br{\!\frac{m_s}{\mCL}\!}$ 
(for comparison, the minimum energy per baryon at $\bag_{60}$ is $883.6232\,\MeV$ for $m_s=150\,\MeV/c^2$, which is $5.5\%$ higher than that for a massless s-quark). For a massless s-quark, the minimum energy per baryon is simply $4b^{1/4}{\cdot}150\,\MeV$, where $b$ is evaluated for $\bag_{60}$ at the $\mCL$ mass scale.

\section{\label{sec:scalingprop}Scaling property of the equations of stellar structure}

\newcommand{\degr}{\mathsf{n}}
Consider scaled profiles $\mycal{X}(\nu)=\lambda^{\degr_X} \mycal{X}_{\lambda}(\nu_{\lambda})$ 
with $\nu_{\lambda}\equiv\lambda{\cdot}\nu$ and the derivative property 
$\mycal{X}'(\nu)=\lambda^{\degr_X+1}\mycal{X}_{\lambda}'(\nu_{\lambda}) $, applied to all functions and parameters, 
each with their respective scaling indexes $\degr_X$. Substituting these into the equations 
\eqref{eq:structure} {and defining  $\tau\equiv \degr_{\mycal{S}}$}, it can be inferred that the ratios $\mycal{Q}'/\mycal{S}'$, $\mycal{M}'/\mycal{S}'$, 
$\mycal{N}'/\mycal{S}'$ and $\mycal{U}'/\mycal{S}'$ remain form-invariant when $\degr_{\mycal{Q}}=\tau$, $\degr_{\mycal{M}}=\tau$ 
 (as required by the structure of the scalar  $1-{2 \mycal{M}}/{ \mycal{S}}+{ \mycal{Q}^2}/{ \mycal{S}^2}$ in the 
 metric function $\mycal{L}$), $\degr_{\mycal{P}}=-2\tau$,  $\degr_{\mycal{R}}=-2\tau$ 
 (implied then by the structure of the scalars $ \mycal{M}+\mycal{P}\mycal{S}^3-\mycal{Q}^2/\mycal{S}$ and  $\mycal{P}+\mycal{R}$ 
 involving pressure), and  $\degr_{\mycal{N}}=3\tau-1$, $\degr_{\mycal{U}}=0$ (following then from equations for $\mycal{N}'$ and $\mycal{U}'$).  
 Hence, the scaling rules implied for solutions of the structure equations are as follows
\begin{equation*}
\begin{aligned} 
&\nu=\lambda^{-1}\nu_{\lambda},
\quad \mycal{M}=\lambda^{\tau}\mycal{M}_{\lambda}, 
\quad \mycal{S}=\lambda^{\tau}\mycal{S}_{\lambda},
\quad \mycal{N}=\lambda^{3\tau-1}\mycal{N}_{\lambda}, \\
&\mycal{P}=\lambda^{-2\tau}\mycal{P}_{\lambda}, 
\quad \mycal{R}=\lambda^{-2\tau}\mycal{R}_{\lambda}, 
\quad \mycal{Q}=\lambda^{\tau}\mycal{Q}_{\lambda}.
\end{aligned}
\end{equation*}
Dimensional analysis suggests that magnetic field $\mycal{H}$  (or electric field $\mycal{E}$), which 
dimensionally mirrors $\mycal{Q}{\cdot}\mycal{S}^{-2}$, should scale as: 
$$\mycal{H}=\lambda^{-\tau} \mycal{H}_{\lambda},\qquad \mycal{E}=\lambda^{-\tau} \mycal{E}_{\lambda}.$$ 
Furthermore, the left side of the remaining equation $\mycal{W}\mycal{S}'=1$ transforms to 
 $\lambda^{-1-\tau}\mycal{W}_{\lambda}\mycal{S}'$. However, since $\mycal{S}'(\nu)=
 \lambda^{\tau+1}\mycal{S}_{\lambda}'(\nu_{\lambda})$, this equation preserves its form for any $\tau$. 
Thus, due to this invariance, if $X(\nu)$ is a solution, then $\lambda^{-\degr_X}X(\nu_{\lambda}/\lambda)$ is also a valid solution.
The scaling law for $\mycal{Q}$ is consistent with the definition inherent in the equation for $\mycal{Q}'$  assumed in \eqref{eq:structure}.

\subsection{\label{sec:masslessscaling}Massless s-quark and other power law equations of state. Exact scaling with the bag constant} 

In the high density $\nu$ limit of the current model, and particularly as $m_s\to0$ as outlined earlier in \secref{eq:eoslim}, the equation of state  \eqref{eq:eosmain} reduces to a linear one: $\mycal{R}=3\mycal{P}+4b$, where $b$ denotes the reduced bag constant, adjusted in the units adopted during this limiting process. Given that pressure is a Legendre transformation of energy density considered as function of particle density, this linear correspondence between $\mycal{R}$ and $\mycal{P}$ is characteristic of a class of power-law equations of state, described as $\mycal{R}=\alpha\nu^{\gamma}+\delta$ (with $\gamma= 4/3$ here, and $\alpha$ being absorbed by the choice of units), then $\mycal{R}=\frac{1}{\gamma-1}\mycal{P}+\frac{\gamma\delta}{\gamma-1}$. In particular, the linear dependence $\mycal{R}(\mycal{P})$  with $1.22<\gamma<1.32$  accommodates a more reliable $\beta$-stable quark matter model in a quasiparticle description \cite{peshier2000} which extends the simple bag model with $\gamma=4/3$.
 For an arbitrary $\gamma$, the general form could be regarded as homogeneous of degree $\gamma$, provided that the parameter $b$ is rescaled as $b=\lambda^{-2s}b_{\lambda}$. This homogeneity aligns with that of the stellar structure equations only if $s=\gamma/2$. Consequently, for any $\gamma$, the solutions manifest the following scaling symmetry with the reduced bag constant $b$:
\begin{equation*}
\begin{aligned} 
&\frac{\nu}{\tilde{\nu}}{=}\br{\frac{b}{\tilde{b}}}^{\!\!1/\gamma},\quad
\frac{\mycal{N}(\nu)}{\tilde{\mycal{N}}(\tilde{\nu})}{=}\br{\frac{\tilde{b}}{b}}^{3/2-1/\gamma}, \quad
\frac{\mycal{P}(\nu)}{\tilde{\mycal{P}}(\tilde{\nu})}{=}\frac{\mycal{R}(\nu)}{\tilde{\mycal{R}}(\tilde{\nu})}{=}\frac{b}{\tilde{b}}, \\
&\frac{\mycal{M}(\nu)}{\tilde{\mycal{M}}(\tilde{\nu})}{=}\frac{\mycal{Q}(\nu)}{\tilde{\mycal{Q}}(\tilde{\nu})}{=}\frac{\mycal{S}(\nu)}{\tilde{\mycal{S}}(\tilde{\nu})}{=}\sqrt{\frac{\tilde{b}}{b}}, \quad
\frac{\mycal{H}(\nu)}{\tilde{\mycal{H}}(\tilde{\nu})}{=}\sqrt{\frac{b}{\tilde{b}}}.
\end{aligned} 
\end{equation*}
It should be emphasized that these scaling relations are local; they concern not only global characteristics of the maxiumum mass stars, such as their total mass and radius, but describe scaling properties of structure functions describing every point in the stellar interior of arbitrary star considered in this model. 
Regarding the potentially observable global characteristics, it is notable that the scaling laws do not differentiate between models with various $\gamma$, unless 
the behavior of the integrated baryon number $\mycal{N}$ is concerned. From this perspective, the model with $\gamma=2/3$ stands out since the scaling would preserve the baryon number; however, this configuration breaks the general requirement from the speed of sound argument, which stipulates $0<\gamma-1\leqslant1$.

Naturally, some of above scaling laws relevant in scenarios where electromagnetic fields are disregarded, appear already in the context of model with $\gamma=4/3$ for global quantities like the maximum mass star and its radius \cite{witten1984}. The generalized scaling laws given above in the presence of magnetic field were observed in \cite{bratek2023} with the specific form of coupling \eqref{eq:magncoupl} between magnetic field and matter. For this reason, this particular coupling has also been adopted here, in order to see the changes introduced by the s-quark mass. 

As discussed at the beginning of \secref{sec:scalingprop}, these local scaling properties arise due to the compatibility of the gravitational and electromagnetic equations' symmetries with the symmetry for a power-law equation of state. Nevertheless, this consistency could be disrupted for more complex equation of state or a more intricate magnetic field structure than considered here. 
In particular, this disruption may already affect the stellar `mass function', $\mycal{M}(\nu)$, due to its direct dependence on the electromagnetic contribution as seen in \eqref{eq:magncoupl} of the present model. 

\subsection{\label{sec:massive-scal}Massive s-quark. Exact and Approximate scaling with the Bag constant} 

The  presence of the threshold value $b_m$ signifies a qualitative change in the equation of state. Specifically, with massive s-quark, the scaling properties detailed in \secref{sec:masslessscaling} for SQM model  with massless s-quark will be only approximately fulfilled.  
It is clear from \secref{sec:criticalbag} that SQM with a massive s-quark is viable only 
if the parameter $b$ exceeds a critical minimum value
$b_m$.  In view of the definition of $b$ in \secref{sec:criticalbag}, this means that for each value of the  s-quark mass $m_s$ there is a corresponding 
lowest permissible bag constant $\bag_m(m_s)$, below which no solutions are feasible, and conversely,
for each value of the bag constant there is a corresponding highest permissible s-quark mass, beyond which no solutions are feasible.  
 
Finally, it should be stressed that SQM stars with the same $b$ are exactly scale-symmetric with $\bag$, provided that the 
s-quark mass is also rescaled according to \eqref{eq:scalingSqarkMass}.
{}
Scaling with the bag constant will be preserved for solutions with the same parameter $b$, 
since then the reduced equation 
of state is not altered, which requires that the s-quark mass is also appropriately rescaled:
\begin{equation}\label{eq:scalingSqarkMass}\frac{m_s}{\tilde{m_s}}{=}
\br{\frac{\bag}{\tilde{\bag}}}^{\!\!1/4}{, \qquad b{=}{const}.}\end{equation}
Since then $\nu=\tilde{\nu}$ and other dimensionless functions are also equal, $\mycal{X}(\nu)=\tilde{\mycal{X}}(\tilde{\nu})$, one obtains stars with rescaled 
structures like for the model with a massless s-quark, however the s-quark mass must also be rescaled so that the ratio $\bag/m_s^4$ remains unaltered by the change of $\bag$.
{}
Then, the symmetry concerns the entire internal profiles of the stars. Therefore it is sufficient to consider only stars with the same representative bag constant, 
in which case the stars differ by the value of $m_s$. To assess the extent of violation of the scaling property for stars with massive s-quark, one may focus on considering 
solutions that have already been rescaled to some standard bag constant, here  $\bagref$, following the scaling rules established for SQM stars with massless s-quark.

\newcommand{\locbfrac}[2]{({#1}{/}{#2})}
\newcommand{\locfrac}[2]{{#1}{/}{#2}}
\begin{table}
\[\! {\small \begin{tabular}{|@{\;}l@{\,}|@{\;}l@{\!\,}|@{\;}l@{\!\,}|}
\hline
I & II & III \\
\hline 
\hline
   $\locfrac{n_r}{\tilde{n}_r}{=}{\locbfrac{m_s}{\tilde{m}_s}}^{3}$
& $\locfrac{\nu}{\tilde{\nu}}{\sim}{\locbfrac{b}{\tilde{b}}}^{3/4}$  
& $\locfrac{n}{\tilde{n}}{\sim}{\locbfrac{\bag}{\tilde{\bag}}}^{3/4}$
\\
   $\locfrac{p_r}{\tilde{p}_r}{=}{\locbfrac{m_s}{\tilde{m}_s}}^{4}$ 
& $\locfrac{\mycal{P}(\nu)}{\tilde{\mycal{P}}(\tilde{\nu})}{\sim}{\locbfrac{b}{\tilde{b}}}^{1}$ 
& $\locfrac{P}{\tilde{P}}{\sim}{\locbfrac{\bag}{\tilde{\bag}}}^{1}$\\
\hline
\hline
   $\locfrac{r_r}{\tilde{r}_r}{=}{\locbfrac{m_s}{\tilde{m}_s}}^{\!\!-2}$
& $\locfrac{\mycal{S}(\nu)}{\tilde{\mycal{S}}(\tilde{\nu})}{\sim}{\locbfrac{b}{\tilde{b}}}^{\!\!-1/2}$ 
& $\locfrac{r}{\tilde{r}}{\sim}{\locbfrac{\bag}{\tilde{\bag}}}^{\!\!{-}1/2}$ 
\\
   $\locfrac{m_r}{\tilde{m}_r}{=}{\locbfrac{m_s}{\tilde{m}_s}}^{\!\!-2}$ 
& $\locfrac{\!\mycal{M}(\nu)}{\tilde{\!\mycal{M}}(\tilde{\nu})}{\sim}{\locbfrac{b}{\tilde{b}}}^{\!\!-1/2}$ 
& $\locfrac{M}{\tilde{M}}{\sim}{\locbfrac{\bag}{\tilde{\bag}}}^{\!\!-1/2}$
\\
   $\locfrac{h_r}{\tilde{h}_r}{=}{\locbfrac{m_s}{\tilde{m}_s}}^{2}$ 
& $\locfrac{\mycal{H}(\nu)}{\tilde{\mycal{H}}(\tilde{\nu})}{\sim}{\locbfrac{b}{\tilde{b}}}^{1/2}$ 
& $\locfrac{H}{\tilde{H}}{\sim}{\locbfrac{\bag}{\tilde{\bag}}}^{1/2}$ 
\\
\hline
\end{tabular}}
\]
\caption{
Summary of scaling properties of selected quantities for SQM stars with massive and massless s-quarks. {\bf Column I}: Dimensional reference scale parameters (these represent the scaling laws permitted by the symmetries of the gravitational structure equations). {\bf Column II}: Dimensionless structure functions of the reduced system of stellar structure equations. {\bf Column III}: Physical dimensional functions of the stellar structure, calculated as the product of elements from Column I and Column II. Note that the symbol $\sim$ indicates that equality for the model with a massive s-quark is only approximate, improving as $m_s$ decreases. This equality becomes exact for stars with a massless s-quark (in this case, $m_s$ in Column I serves as a reference mass scale, at which the parameter $b$ of Column II assumes the role of the bag constant $\bag$ in Column III). For higher $m_s$ values, the profile functions in Column II become dependent on $b$, as the equation of state deviates from the scaling property typical of the power-law equation of state $\mycal{R}=\alpha \nu^{\gamma}+\delta$ (with $\gamma=4/3$ specifically for SQM stars with a massless s-quark). Note also that SQM stars sharing the same $b$ value exhibit exact scale-symmetry with $\bag$ (or with $m_s^{1/4}$), and this symmetry concerns the entire internal profiles of the stars.
}
\end{table}

\section{Results}

The idea of using the parametrization defined in \eqref{eq:parametric}, is simple, although its realization is demanding.
The functions $\Pi(\nu)$ and $\Epsilon(\nu)$ are not known in explicit form. They are known only in parametric form, meaning that $\Pi(x)$, $\Epsilon(x)$, and $\nu(x)$ are known as algebraic functions of the auxiliary variable $x$.
Therefore, for the numerical integration of the system \eqref{eq:structure}, which uses $\nu$ as the independent integration variable, the function  $\Chi(\nu)=x$, which is the inverse of  $\nu(x)$, needs to be determined on the run at arbitrary value of $\nu$, together with its derivatives. 
The function $\Chi(\nu)$ could be obtained straightforwardly through appropriate interpolation of the transposed matrix representing the discretized function $\nu(x)$ and their derivatives. For the purposes of this study, however, a specially written, quickly convergent numerical root-finding procedure was devised. 

Given a particular value of $\nu=\nu_o$ one starts with  a learned first guess for $x\in(0,x_m)$ and, knowing the analytic form of $\nu(x)$ and a suitable number of its derivatives, refines it iteratively with an algorithm akin to Newton-Raphson method of a higher order involving $\nu'(x)$ and $\nu''(x)$. The success of applying this method is subject to a good guess of the starting approximation of $x_o$, which is central to this method.

The approximation is found by combining the asymptotic expansion $x(\nu)$ at $x=0$ already found above, \eqref{eq:inverseseries1}, and a series expansion at $x_m$. The latter is more critical for this method since the derivatives of $\nu$ diverge at $x_m$.  At this point $d\approx d_m$ is bound from $0$, while $s\to0$.  Therefore, one can expand the formula $\nu=1/(d^{2/3}-s^{2/3})^{3/2}$  with respect to $s$ obtaining $\nu=d^{-1}(1+(3/2)(s/d)^{2/3}+\dots)$  and consider the limit $d\to d_m$. Then, substituting the series expansion $s(x)=s'_m(x-x_m)+\dots $, one can infer that  in the leading order $$x\sim x_m-\gamma_m\br{{\nu}/{\nu_m}-1}^{3/2}\quad {\rm as} \quad  \nu\to\nu_m,
 $$ where $ \gamma_m=-(d_m/s'_m)(2/3)^{3/2}$, see \tabref{tab:param}.   
Once the method has converged to the root $x_o$,  the corresponding values of the remaining 
functions and their derivatives with respect to $\nu$ are determined with the required accuracy at 
every integration step  by substituting $x=X(\nu)$ to all needed expressions (higher derivatives are also required for the series expansions of the solutions used by the numerical integrator to start the integration from the singular point of the system \eqref{eq:structure} and to determine the optimum initial step size).

Owing to the extensive numerical computations required in the numerical integration process, 
appropriately designed algebraic formulas and careful control of the precision of the results is essential . 
The computations were carried out with due care, focusing on controlling the precision of the functions and the step size. 
{A high-precision adaptive integration algorithm} 
 and {all}{the} auxiliary functions {described above} were 
{implemented} in 
the {\it Wolfram Mathematica} programming language \cite{bib:mathematica}, 
independently of the {system's} built-in numerical library.

\begin{figure}[htb]
\centering
\includegraphics[trim=0 0 0 0,clip,width=1\columnwidth]{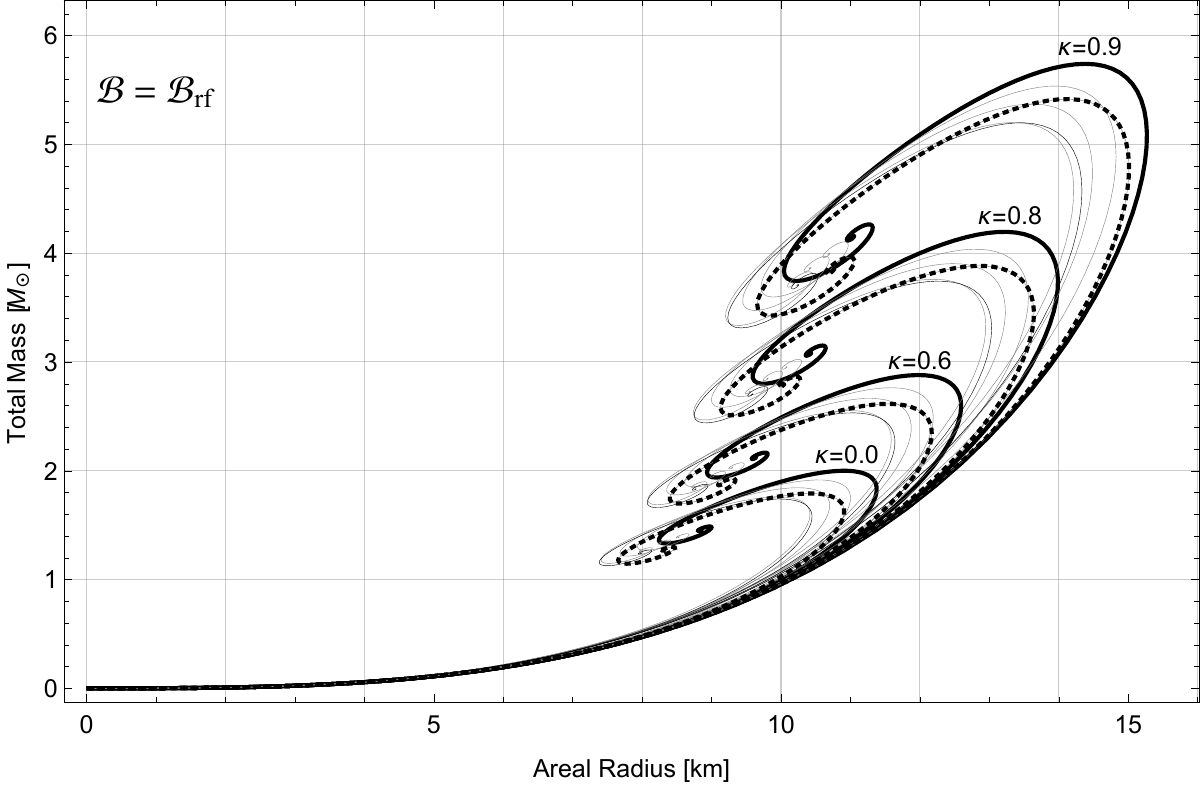}
\includegraphics[trim=0 0 0 0,clip,width=1\columnwidth]{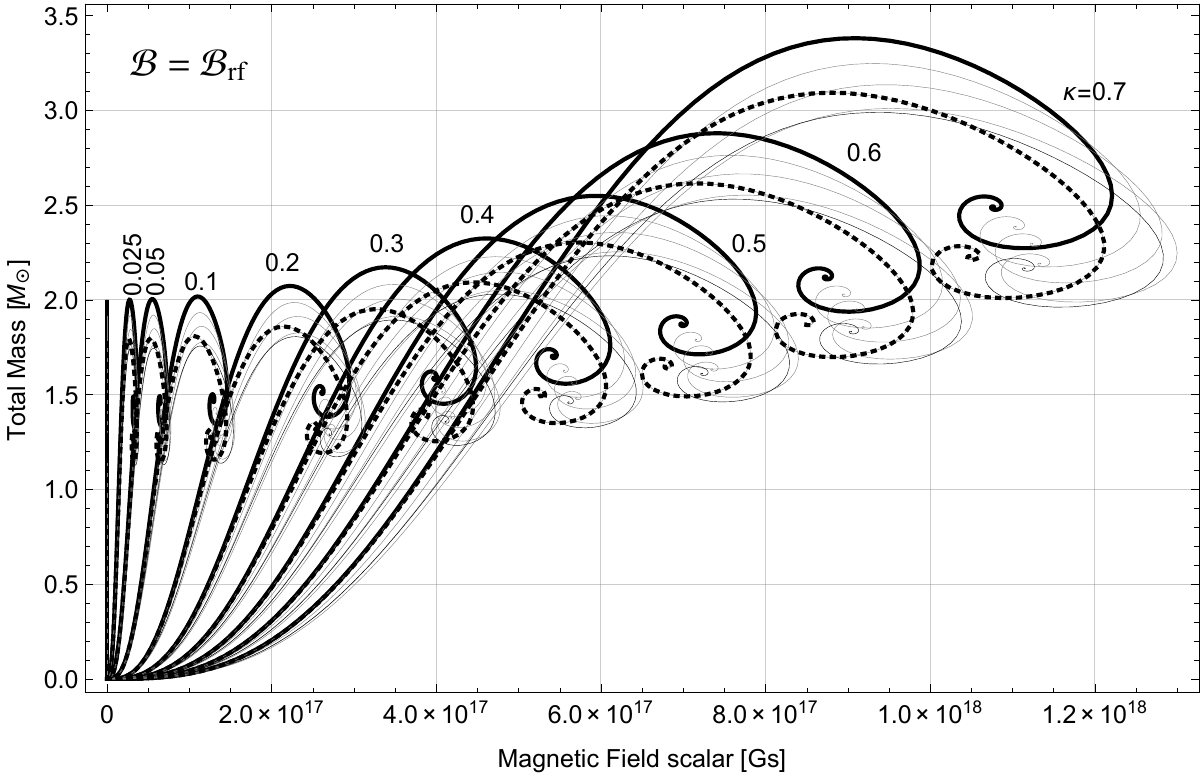}
\caption{\label{fig:RM-diagram}
The characteristics of magnetized SQM stars with a massive s-quark rescaled to the reference bag constant value $\bag{=}\bagref$. The Total Mass vs. Radius diagram [{\it top panel}] and the Surface Magnetic Field vs. Total Mass diagram [{\it bottom panel}]. Each diagram displays example families of lines, each family corresponding to a different $\kappa$ ratio parameter. Within each family, six example lines represent different s-quark masses ($m_s$): {\it thick black lines} represent symmetric SQM stars (massless s-quark, $m_s{=}0$, $b{=}\infty$) with a maximum mass of $2\Msun$; 
{\it thick black dotted lines} represent stars with the highest s-quark mass of ${\sim}334\,\MeV{/}c^{2}$ possible at $\bag{=}\bagref$ (in this case, $b{=}b_m$);
{\it thin black lines} corresponding to $m_s{\sim}269\,\MeV{/}c^2$ are continuous deformations of the line at $\kappa{=}0$ that include a star at the infimum of $1.74\,\Msun$ for maximum mass stars without a magnetic field; {\it thin gray lines} correspond to example s-quark masses of $m_s{\sim}100,\,150$, and $222\,\MeV{/}c^{2}$.
 }
\end{figure}

\begin{figure*}
\centering
\begin{tabular}{l@{\hspace{0.05\columnwidth}}r}
\includegraphics[trim=0 0 0 0,clip,width=1\columnwidth]{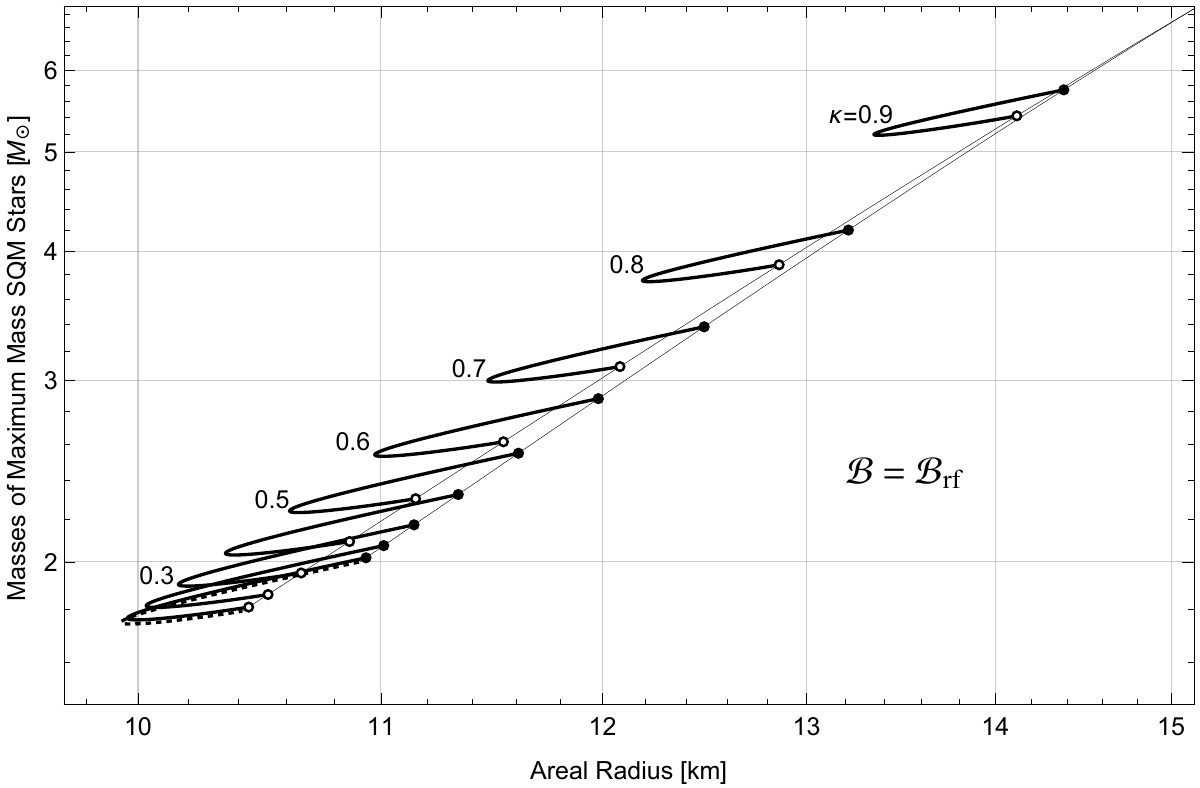} &
\includegraphics[trim=0 0 0 0,clip,width=1\columnwidth]{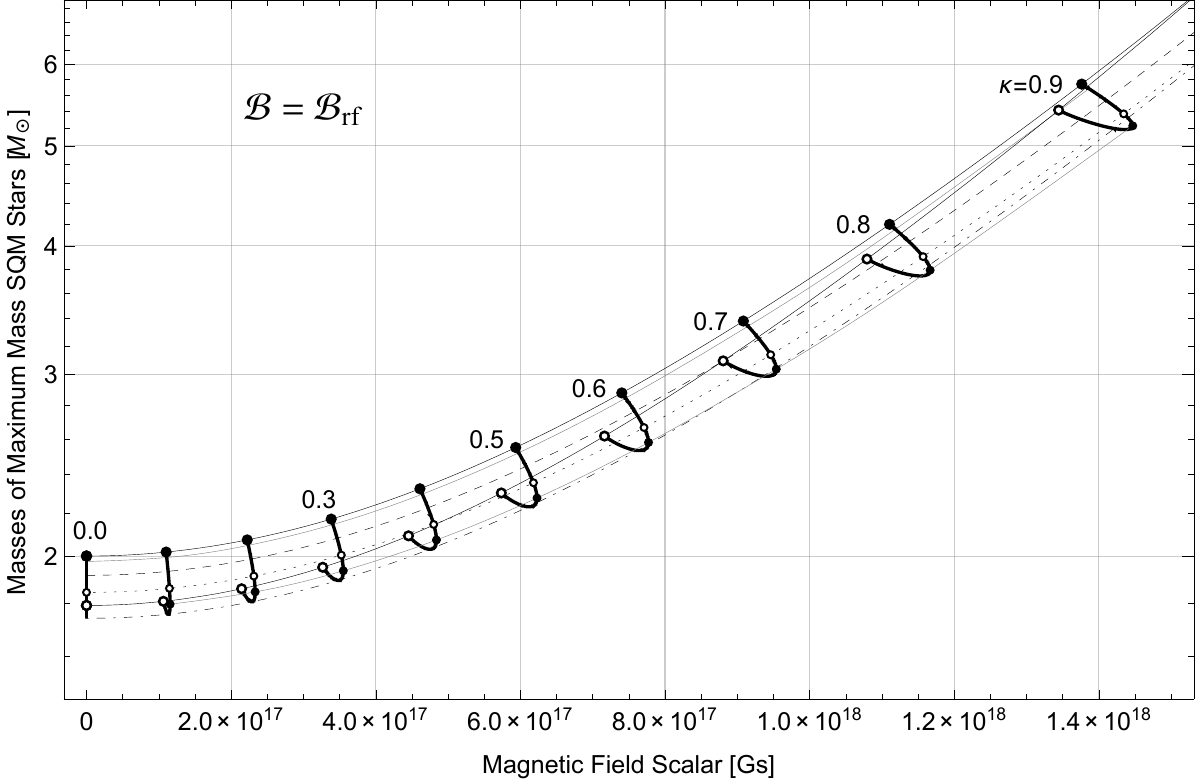}
\end{tabular}
\caption{\label{fig:maxMassSol}
Diagrams of Maximum-mass magnetized SQM stars with massive s-quark, shown for different $\kappa$ ratios and different s-quark masses on the {\it Radius vs. Mass}  Log-Log plane and on the {\it Magnetic Field Scalar vs. Mass} Linear-Log plane, all scaled to a reference bag constant $\bagref$.
{}\\
{\bf Both panels:}
The {\it thin line through filled black circles} -- represents a family of symmetric-SQM stars (with massless s-quark), parameterized with the $\kappa$ ratio varying continuously along this line (data for row S of table  \tabref{tab:fitparam});
The {\it thin line through empty black circles} -- represents a family of stars with the highest s-quark mass of $334.4\,\MeV{/}c^{2}$ permissible at $\bagref$, parameterized with the $\kappa$ ratio varying continuously along this line (row P of table  \tabref{tab:fitparam});
The {\it thick black U-shaped solid lines} -- each represents the entire spectrum of s-quark masses  at the indicated discrete values of the $\kappa$ ratio, starting from the massless s-quark limit ({\it filled black circle}) and ending at the highest s-quark mass limit ({\it empty black circle}).\\ 
{\bf Left panel:}
The {\it dashed U-shaped line} represents stars in the $\kappa=0$ limit (without magnetic fields). 
{}\\
{\bf Right panel:} 
The {\it thin gray solid line} are stars with s-quark masses of about $50\,\MeV{/}c^{2}$  at $\bag=\bagref$;
The {\it thin dashed line} are stars with s-quark mass of $100\,\MeV{/}c^{2}$ (data for row A of table \tabref{tab:fitparam});
The {\it dotted line through smaller empty black circles} indicates stars with s-quark mass of $m_s=150\,\MeV{/}c^2$ (data for row B in  \tabref{tab:fitparam}); 
The {\it thin solid gray line through smaller filled black circles} indicates stars with the highest value of magnetic field at a given value of $\kappa$ (with s-quark masses varying a little from $201\,\MeV{/}c^2$ at $\kappa=0.1$ to $204\,\MeV{/}c^2$  at $\kappa=0.9$);
The {\it thin dotted-dashed line} are stars with s-quark mass of $269\,\MeV{/}c^{2}$ (data for row D of table  \tabref{tab:fitparam}) -- this line includes the lowest mass star at $\kappa=0$.
}
\end{figure*}

The result of the numerical integration is a set of data representing full internal structure of stars. Different stars  are tagged with 
different triples ($\kappa,\m_s,\bag$), with $0<\kappa<1$ for magnetization degree $\kappa$ and $b_m\leqslant b(\bag,m_s)<\infty$ 
for $b$. Additionally, the data are extended by the limiting solutions without magnetic field ($\kappa=0$) or those 
with massless s-quark ($m_s=0$).
As described in \secref{sec:massive-scal}, solutions with the same $b$ can be mapped to each other with the help of scaling transformation 
rules for the structure functions and parameters $m_s$ and $\bag$, at the same time the parameter $\kappa$ is not changed. Accordingly, 
any solution $ (\tilde{\bag},\tilde{m_s},\tilde{\kappa})$ will be rescaled to a reference bag constant $\bagref$: 
$$(\tilde{\bag},\tilde{m_s},\tilde{\kappa})\quad\to\quad (\bag,m_s,\kappa)= (\bagref,\tilde{m_s}\sqrt[4]{\sfrac{\bagref}{\tilde{\bag}}},\tilde{\kappa}).$$ 
The rescaled solutions are shown in figure \figref{fig:RM-diagram} on the mass-radius diagram and the mass-field diagram.
{}
{}
In particular, at this reference value, interested are maximum mass stars.
It is a priori difficult to predict which combination of  $(\tilde{\bag},\tilde{m_s},\tilde{\kappa})$ parameters to chose such that one obtains precisely the maximum mass star. However, as seen in  \figref{fig:RM-diagram}, the maximum-mass stars are local maxima on the diagram lines which are concave curves. It is enough to write an algorithm that first automatically identifies three points in a small neighborhood of the unknown maximum (with two of the points on both sides of the third point with the highest mass), then the algorithm  generates a sufficient number of new star models, then it determines the local maximum by finding the best qubic polynomial fit, and finally the unique maximum is found from simple algebraic expressions (without the need of Cardano formulas for the roots) and the values of other structure functions can be then determined. Such found parameters of maximum mass solutions locate along lines on the diagrams as presented in \figref{fig:maxMassSol}.     
{}
{}
The dependence of maximum mass on the magnetic field, as seen in this figure, seems quite complicated. It is 
characteristic that when $m_s$ increases from zero, then the star mass initially decreases attaining its lower bound and then it starts to increase as $m_s$ increases toward its upper bound determined by the threshold value $b_m$ of the \rem{compactness} parameter $b$.  
Strong magnetic field changes this behaviour quantitatively but not qualitatively.  Similarly, keeping $\kappa$ constant the surface magnetic initially 
increases with $m_s$ to its maximum value and then starts to decrease while $m_s$ still grows. This behaviour may seem fairly complicated.  
It would be helpful to have a single formula that describes the value of mass of the maximum mass star 
with a given magnetic field strength on its surface for any values of the parameters $\bag$ and $m_s$. Such a phenomenological formula can be 
found as shown below.

\subsection{Phenomenological scaling formula}

The mass integral \eqref{eq:totalmass}  is an even functional of the magnetic field scalar. This can be seen from a formal generalized power series 
expansion in the variable $\delta\nu=\nu-\nu_c$ of all independent functions of baryon density $\nu$ satisfying the system of differential equations \eqref{eq:structure}.  These series were also utilized to propagate the initial data from the singularity point of these equations at the star center with central density $\nu_c$ to a regular neighboring point $\nu=\nu_c+\delta\nu$ tractable by the numerical integrator. 

The series coefficients involve only even powers of $\kappa$ and are functions dependent on $\nu_c$ and the {parameter} $b$. 
This suggests that upon eliminating $\delta\nu$ using the inverse series method, both the mass integral and the magnetic scalar, as evaluated at the star 
surface (where the series for pressure vanishes), can be regarded as functions of each other at the star surface. 
Therefore, for a given value of $b$, the mass $\mathcal{M}_{\star}(b,\mathcal{H}_{\star})$ calculated at some bag value $\bag_{\star}$ will be a power 
series involving only even powers of the magnetic scalar $\mathcal{H}_{\star}$, and this series will converge to some limiting value 
$\mathcal{M}_{\star}(b,0)$ in the absence of the field. 

Furthermore, the natural scale of the magnetic field  in the adopted units, as established in \eqref{eq:hunit}, depends on a particular power of the s-quark mass parameter $m_s$. Since the $b$ ratio, as defined in \eqref{eq:bdef}, is set to be a constant, varying the scale of mass $m_s$ is equivalent to varying the bag constant $\bag$ so as the ratio $\bag/m_s^4$ remains constant. 
Therefore, for a given value of $b$, the ratio $$\mathcal{H}/\sqrt{8\pi\bag},$$ which is dimensionless and involves  in the denominator the natural measure of magnetic field strength defined similarly to \eqref{eq:magbagunit}, should be used as a scale-independent magnetic variable in the generalized formulas. 
This ratio appears naturally in the scale-invariant symmetric--SQM model which can be considered as a limiting one  as $b\to\infty$. Moreover, the same reasoning concerning the scales as applied to the magnetic field can be applied to the star mass, leading to a scale-invariant product $$\mathcal{M}\sqrt{\bag}.$$ 

\noindent
In conclusion to the above remarks, it can be stated that the continuation of the scaling property of the symmetric model to the entire spectrum of $b$ parameters should proceed through postulating the following analytic function:
$$\mathcal{M}(b,\mathcal{H})=\mathcal{M}_{\star}(b,0)\sqrt{\frac{\bag_{\star}}{\bag}}\br{1+\sum\limits_{k=1}^{\infty}\alpha_k(b){\frac{\mathcal{H}^{2k}}{\bag^k}}}.$$
Here, the series coefficients $\alpha_k$ are dimensionless functions of $b$ only. The initial terms  of this series can, in principle, be found for the symmetric-SQM star with a simple equation of state (for example, it is possible to obtain this way the mass-radius relation in the absence of the magnetic field; however, this approach is feasible only for star radii much lower than for the maximum mass stars, otherwise the number of required expansion terms quickly increases). The problem becomes considerably more difficult for the present equation of state. Because of this difficulty, a simple formula that reconstructs the numerical data for the maximum mass stars at an arbitrary field strength and a given $b$ value, and then tries to interpolate between these fitting curves, can be considered instead.  This approach is realized below. 

The following function is taken as the fitting model to reconstruct the lines of constant $b$ on the  $\mathcal{H}$-$\mathcal{M}$ plane 
such as displayed in the right panel of \figref{fig:maxMassSol}: 
\newcommand{\cfC}{\gamma_b}
\newcommand{\cfB}{\beta_b}
\newcommand{\cfA}{\alpha_b}
\newcommand{\cfX}{\chi_b}
\newcommand{\cfM}{\mu_b}
\newcommand{\ratx}{\frac{ \chi^2}{\cfX^2}}
{}
\begin{equation}\label{eq:fitfmodel}f_b(\chi)=\cfM\br{\!1+\cfC\ratx\frac{\br{1+\cfB\ratx}}{\br{1-\ratx}^{\cfA}}\!}
,\quad \chi=
\frac{\mathcal{H}[\Gs]}{\sqrt{8\pi\bag[\frac{\erg}{\cm^{3}}]}}.\end{equation}
For $\alpha_b=0$, this function results in an even polynomial of degree four in the magnetic scalar $\mathcal{H}$, which accurately reconstructs the target curve for small $\mathcal{H}$ with effectively $3$ parameters. As $\mathcal{H}$ increases, the required polynomial degree and thus the number of free parameters grow, which is undesirable. A straightforward method to introduce an infinite number of terms without affecting the lowest degree terms and ensuring an almost linear behavior on a Log-Log plot (indicating a power law behavior) for larger $\mathcal{H}$ is to use a generalized rational function. In the simplest case, this involves a monomial in the denominator raised to an arbitrary positive power.  Since the mass quickly grows for larger $\mathcal{H}$, this behavior can be effectively mimicked when the root of the monomial is almost reached, which is possible with the minus sign in front of $\chi^2$ in the monomial. This simple correction significantly
improved the accuracy of the polynomial fit. This simple profile proves to be quite suitable in the considered case, performing well with a small number of free parameters for any $b$.  The result for the best fit parameters at the reference bag constant $\bag=\bagref$ is displayed in table \tabref{tab:fitparam} for several representative curves with different $b$ values.

\begin{table*}
\!\!\!\begin{tabular}{|@{\;}c@{\;}|@{\,}c@{\,}|@{\,}c@{\,}|@{\;}c@{\;}||@{\;}c@{\;}|@{\;}c@{\;}|@{\;}c@{\;}|@{\;}c@{\;}|@{\;}c@{\;}|@{\;}c@{\,}||@{\;}c@{\;}|}
\hline
& I & II & III & 0 & 1 & 2 & 3 & 4 & 5 & E \\
\hline
& $\bag$ & $m_s$ & $b/b_m$ & $m_i$ & $\cfM$ & $\cfX$ & $\cfC$ & $\cfB$ & $\cfA$ & $\Delta_{\rm max}$\\
& $[\sfrac{\MeV}{\fm^3}]$ & $[\sfrac{\MeV}{c^2}]$ & $--$ & $[\sfrac{\MeV}{c^2}]$ & $[\Msun]$ & $--$ & $--$ & $--$ & $--$ & $--$ \\
\hline\hline
S & $58.0$ & $0.00$ & $\infty$ & $0.000000$ & $2.000280(84)$ & $1.218520(33)$ & $2.54516(48)$ & $0.46204(28)$ & $0.084601(59)$ & $0.00019$\\
A & $100.$ & $115.$ & $123.7090$ & $100.2767$ & $1.916270(41)$ & $1.250190(57)$ & $2.56135(45)$ & $0.47702(28)$ & $0.086513(86)$ & $0.00030$\\
B & $4.94$ & $81.1$ & $24.70792$ & $150.0000$ & $1.84488(11)$ & $1.271670(33)$ & $2.56627(82)$ & $0.50157(62)$ & $0.08523(10)$ & $0.00056$\\
C & $2.46$ & $101.$ & $5.114960$ & $222.3769$ & $1.76121(15)$ & $1.285420(46)$ & $2.6065(11)$ & $0.53433(86)$ & $0.08694(14)$ & $0.00075$\\
D & $53.7$ & $264.$ & $2.391945$ & $268.9134$ & $1.74177(16)$ & $1.268750(50)$ & $2.6500(12)$ & $0.55552(90)$ & $0.08656(15)$ & $0.00079$\\
P & $33.6$ & $292.$ & $1.000001$ & $334.4260$ & $1.791820(97)$ & $1.216570(37)$ & $2.80436(76)$ & $0.57504(52)$ & $0.074141(92)$ & $0.00046$\\
\hline
 \end{tabular}
 \caption{\label{tab:fitparam}Best fit parameters for the model curve \eqref{eq:fitfmodel}, which describes the masses of maximum mass stars as a function of surface magnetic field at a given value of the \rem{compactness} parameter $b$. The parameters were determined at the reference bag value $\bagref$ for six representative values of $b$. \\
Column I: Exact value of bag constant for a given data set of SQM stars. \\
Column II: Assumed exact value of s-quark mass for the data set. \\
Column III: The corresponding $b$ value relative to the absolute threshold value $b_m$. \\
Column 0: The corresponding s-quark mass at the $\bag=\bagref$ scale. \\
Column 1: Optimum mass parameter (the predicted star mass at vanishing magnetic field). \\
Columns 2, 3, 4, and 5: Remaining dimensionless parameters. \\
Column E: Here, $\Delta_{\rm max}=|(\mathcal{M}_p/\mathcal{M}_o)-1|$ represents the maximum relative error for SQM star mass 
prediction with an arbitrary surface magnetic field, where $\mathcal{M}_o$ is the observed numerical value of mass and \ $\mathcal{M}_p$ is the mass predicted by the best fit model curve.\\
Rows S, A, B, C, D, and P: correspond to the six considered data sets for six assumed values of parameter $b$. Specifically, row S are the maximum mass symmetric--SQM stars parameterized with the value of surface magnetic field, similarly row P correspond to threshold s-quark mass, while row D includes the star with the lowest possible mass of $1.74\,\Msun$    attained in the absence of magnetic field at $\bag=\bagref$.
}
\end{table*}

 Each data set consisted of at least $99$ `measurement points' of maximum mass stars with different magnetic fields and the same s-quark mass, with suitably adjusted weights to account for small inhomogeneity of the distribution on the $\mathcal{H}$ axis.  
 For all data sets, the maximum relative error is quite small, less than $0.0008$ in the worst case (as shown in table \tabref{tab:fitparam}), while  the corrected coefficient of determination is virtually $1$ in each case (not shown in the table). This suggests that the rational curve model is very accurate in predicting the obtained data for all the s-quark mass values admissible at $\bag=\bagref$. 

\begin{figure}[htb]
\centering
\includegraphics[trim=0 0 0 0,clip,width=1\columnwidth]{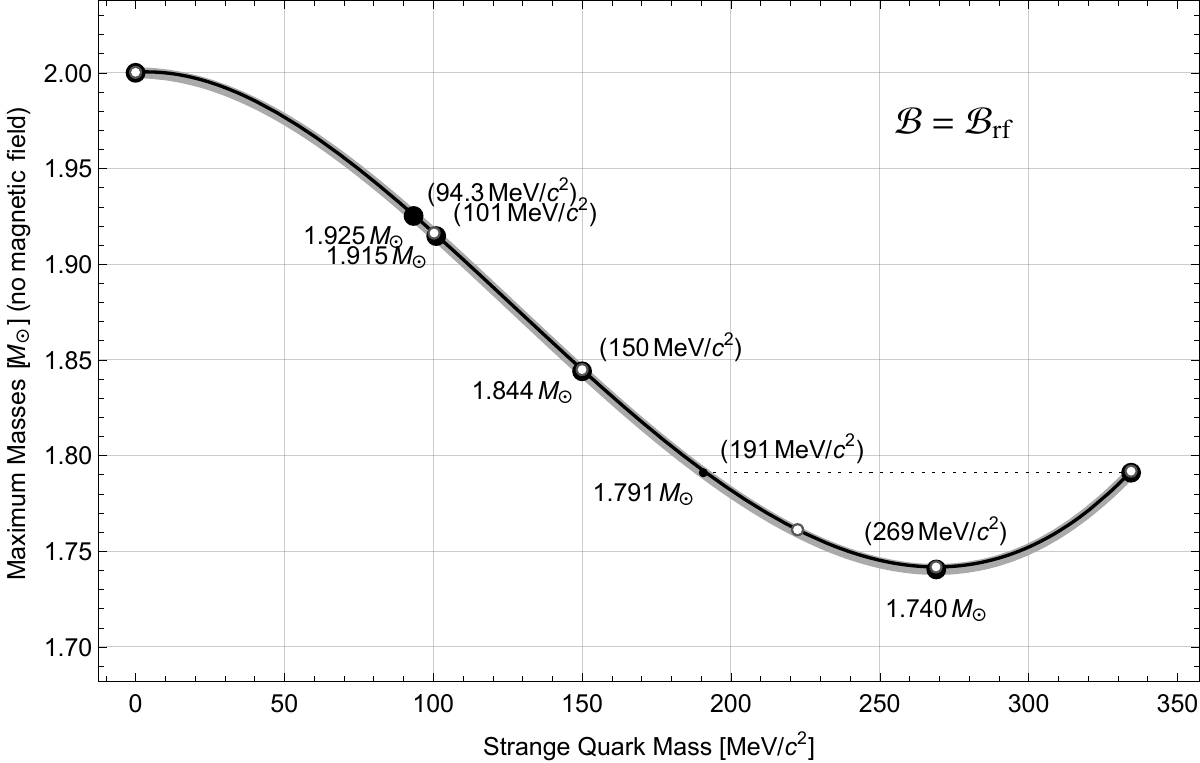}
\caption{\label{fig:mass-vs-sqmass}
{Maximum masses of SQM stars as a function of} the s-quark mass 
{in the absence of a} magnetic field.
{The maximum masses increase for  large enough s-quark masses, which is contrary to the intuition of 
softening the SQM equation of state by the s-quark mass.}
The {\it solid gray line} {is a line plot through 192 points of} real numerical data for the masses of maximum mass stars for different values 
of the s-quark mass at the same reference bag constant $\bagref$ (the symmetric SQM star has a precise mass 
of $2\Msun$). The {\it large black solid circles} points {with the indicated values of the s-quark mass} are 
obtained from an interpolation function through the real 
data points.   The data reach a minimum 
of $1.740\,\Msun$ at {$269.0\,\MeVcc$} and then
terminate at $1.791\,\Msun$ at an s-quark mass of {$334.4\,\MeVcc$}, corresponding to the threshold parameter $b_m$. 
The {\it black solid line} is the Lagrange interpolating polynomial through optimum mass points with coordinates as indicated in 
the columns 0 and 1 of \tabref{tab:fitparam}, marked with the white dots representing $6$ nodal points of the polynomial 
(the line lies slightly above the real data points for the reason explained in the text). Two maximum mass stars of the same 
mass are possible for {$m_s{>}191\,\MeVcc$} within the s-quark mas interval  indicated by the horizontal dotted straight line. 
{The maximum mass values are consistent with those given by \citet{zdunik2000} for $m_s\leqslant200\,\MeV\!/\!c^2$.}
}
\end{figure}

The behaviour of the most important parameter  $\cfM$, which describes masses in units of the solar mass of the 
maximum mass stars in the absence of the magnetic field, is shown as a function of the s-quark mass in 
\figref{fig:mass-vs-sqmass} {shown with a solid black line}. 
{}
{}
This line is the Lagrange interpolating polynomial through the tabulated values displayed in columns 0 and 1 of 
\tabref{tab:fitparam}. This line lies slightly above the actual masses of the maximum mass stars determined from 
the numerical integration data shown in \figref{fig:mass-vs-sqmass} {with a solid gray line}. This is because the parameters in 
columns 0 and 1 of \tabref{tab:fitparam} are 
already optimized as the result of a fitting procedure based on maximum mass data in the presence of extremely strong magnetic fields. 
However, it should be clear that the overlap with the data would be much better at the vanishing field if stars with weaker 
fields were taken into account while preparing the parameters in columns 0 and 1 of  table \tabref{tab:fitparam}.  

\begin{figure}[htb]
\centering
\includegraphics[trim=0 0 0 0,clip,width=1\columnwidth]{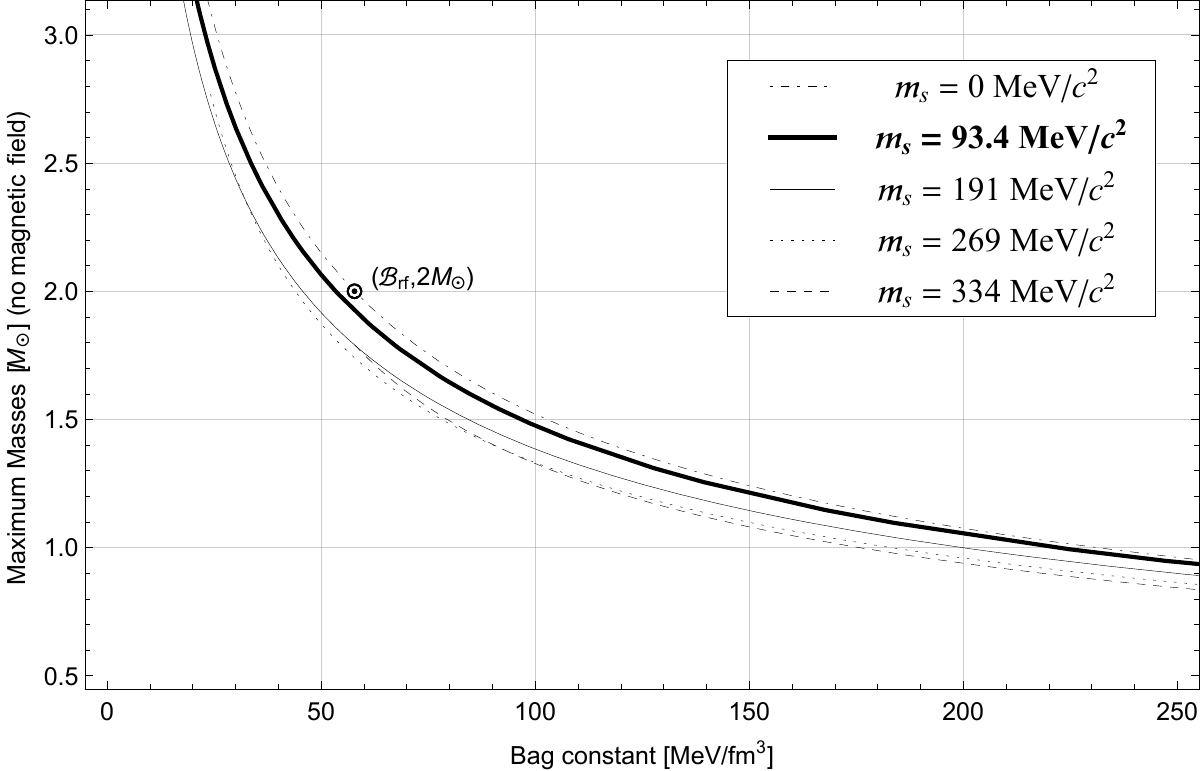}
\caption{\label{fig:mass-vs-bag}Maximum masses of SQM stars as a function of the bag constant 
in the absence of a magnetic field, for various s-quark masses. The lines terminate at their respective 
 critical bag constant values.}
\end{figure}

 The curve in \figref{fig:mass-vs-sqmass} has a local minimum, in the  neighbourhood of which, between $191\,\MeV/c^2$  and $334\,\MeV/c^2$,  
 two maximum mass stars of the same mass are possible with different s-quark masses.
In figure \figref{fig:mass-vs-bag} the maximum-mass diagrams are represented for variable bag constant at several representive 
values of the s-quark mass.
The same interpolation described above and illustrated in \figref{fig:mass-vs-sqmass} for the case without a magnetic field can be repeated for any value of the magnetic scalar $\mathcal{H}$ in the presence of the field. For this purpose it is enough to interpolate between functions \eqref{eq:fitfmodel}  with different fixed values of $b$. This is realized by the formula interpolating between the best-fitting profiles obtained earlier:
 \begin{equation}\label{eq:interpolfit}
\!\!\!\!\mathcal{M}(\!\bag,\!m_s,\!\mathcal{H}){=}\sqrt{\!\frac{\bagref}{\bag}\!}{\;\cdot}\!\sum_{i}\!L_i\!\!\br{\!\!m_s \sqrt[4]{\frac{\bagref}{\bag}}}\!{\cdot} f_{b_i}\!\!\br{\!\!\frac{\mathcal{H}}{\magref}\sqrt{\!\frac{\bagref}{\bag}}}\!.\end{equation}
The summation in \eqref{eq:interpolfit} is taken over the fitting parameters listed in column $0$ of table \tabref{tab:fitparam}. In this equation, 
$L_i(m)$ are Lagrange's fundamental interpolating polynomials with nodal masses  $m_i$ determined at a given $b_i$ value from the equation 
$b(\bagref,m_i)=b_i$, see \eqref{eq:bdef}. To check the performance of this formula, it was applied to all maximum mass solutions, 
especially to those in figure \figref{fig:maxMassSol} which lie on the u-shaped families of curves of constant $\kappa$ and variable $b$ 
values, thus not included in the fitting procedure so far that took into account only maximum mass stars from the family of curves of constant 
$b$ and variable $\kappa$. 

\begin{figure}[htb]
\centering
\includegraphics[trim=0 0 0 0,clip,width=1\columnwidth]{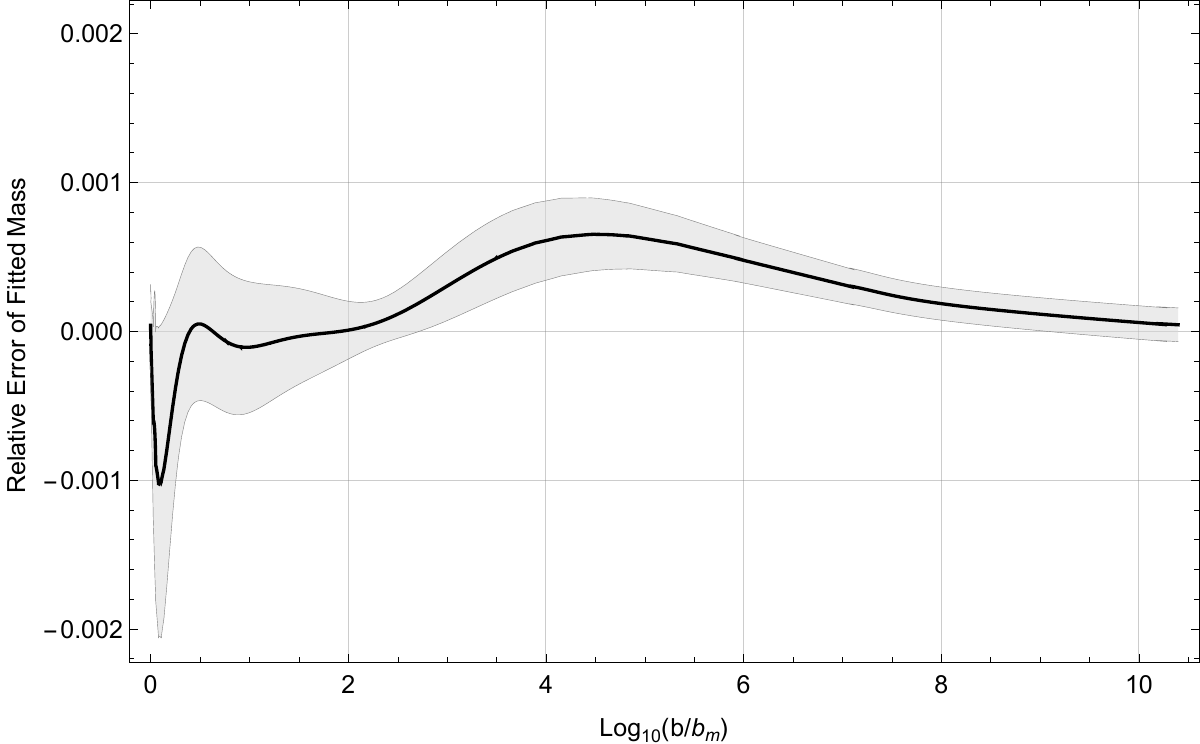}
\caption{\label{fig:rel-err}
Variability and confidence in the stellar mass relative error of the interpolating fit model \eqref{eq:interpolfit} as a function of the 
fraction $b/b_m$ of the \rem{compactness} parameter  $b$ and its threshold value $b_m$. The {\it solid black line} represents the mean 
relative error computed for a set of maximum-mass star numerical solutions with different  $\kappa$ values and the same $b$.
 The {\it shaded region} indicates the corresponding standard deviation of the relative error.}
\end{figure}

As a dimensionless measure of 
discrepancy between the real mass and the predicted mass with the formula \eqref{eq:interpolfit}, the relative difference between these two quantities was calculated for all solutions. These numbers were then partitioned with respect to the same value of $b$ irrespectively of the value of the magnetic field, and the mean value and standard deviation for each $b$ over all $\kappa$ were computed. 
The result is shown in figure \eqref{fig:rel-err}.{}
It is seen that the mean value of the relative error is less or much less than $0.001$ for most solutions, with the deviation also less than 
$0.001$. Therefore, formula \eqref{eq:interpolfit} proves to be reliable for determining the masses of maximum mass stars for all pairs $(\bag,m_s)$ at any value of surface magnetic field in the model investigated in this paper.

\subsection{Compressibility measure in the symmetric and in the s-quark mass--dominated regime.}

The curve in \figref{fig:mass-vs-sqmass}, which predicts lower maximum masses at lower strange quark masses, 
exhibits behavior characteristic of matter with a softened equation of state compared to symmetric SQM.
In contrast, the curve bends upward at higher s-quark masses, indicating that the maximum mass of the star increases with increasing s-quark 
mass.     This strange behavior for large $m_s$ is characteristic of the $m_s$ dominated regime of the equation of state close to the limit $s=0$, 
which cannot be reconstructed from the low $m_s$ limit (see \figref{eq:EPidef} and the discussion in \secref{eq:eoslim}).
 
As a measure of stiffness, the adiabatic index $\gamma=\frac{\rho}{p}\frac{\ud{p}}{\ud{\rho}}$ is commonly used. 
In particular, it describes the response of pressure to changes in energy density. 
This quantity can be derived by the requirement that, aside from involving the first differentials of $P$ and $R$, 
it remains dimensionless and invariant under independent rescaling of both pressure and energy density. 
This scaling invariance is particularly relevant in the context of symmetric SQM stars. 
 However, for SQM, the energy density remains finite at vanishing pressure, which means that $\gamma$ tends to infinity at the 
 star's surface, where pressure is defined to be zero. 
 
 To circumvent this issue, one can instead examine the behavior of the 
 speed of sound expressed in units of the speed of light. This quantity remains invariant under simultaneous rescaling of both 
 pressure and energy density, which is sufficient in the context of gravitating symmetric SQM when the enthalpy form $R+P$ plays a role,
  allowing only identical rescaling factors for $R$ and $P$.  
  
 The speed of sound curve  may be represented on a plane as
 a path $(\mycal{P}(\nu),\mycal{P}'(\nu)/\mycal{R}'(\nu))$,
 thus completely removing the dependence on $\nu$. Therefore, the essential dependence of the model on $m_s$ 
 at a given $\bag$ value comes solely through $b$.
By setting the bag constant equal to a reference value $\bagref$, a change in $b$ corresponds to a change in $m_s$ within the model. 
Other solutions are then obtained by simultaneously rescaling both $m_s$ and $\bag$ while keeping the ratio $\bag/m_s^4$ constant.  
  
  Applied to the present model, the parametric description of the curve 
 follows from differentiating the relations in \eqref{eq:zwiazkifunkcji}:
 $$\frac{v_s^2}{c^2}\equiv\frac{P'}{R'}=\frac{1}{3}-\frac{4}{9}\frac{\br{s\nu}'}{R'\sqrt{1+(s\nu)^{2/3}}}.$$ 
 Remarkably, this expression is universal:  being  independent both 
 of $\bag$ and $m_s$, it is also re-parametrization invariant -- the prime denotes differentiation with respect to any independent variable. 
 Additionally, the `conformal' propagation velocity contribution of $c/\sqrt{3}$ is characteristic of the limiting  symmetric SQM model. 
 Surprisingly, however, the correction term never 
 vanishes and cannot be simply disregarded; it remains also in the $m_s=0$ case. 
  
  However, the physical content depends on which part of the 
 curve the physical variable covers. In particular, by taking $P$ as the independent variable,
   the curve can be rewritten in the form obtained by formally setting $P'\equiv 1$ 
 and $R'\equiv\ud{R}/{\ud{P}}=c^2/v_s^2$, and solving the original formula for the velocity, 
  resulting in
  $$\frac{v_s^2}{c^2}=\frac{\ud{P}}{\ud{R}}=\frac{1}{3}
 \frac{
 \sqrt{1+(s\nu)^{2/3}}
 }{
 \sqrt{1+(s\nu)^{2/3}}+\frac{4}{9}
 \frac{\ud{\br{s\nu}}}{\ud{P}}
 }, \quad P\geqslant 0,$$ where all quantities are regarded as functions of $P$. 
 In this case, the dependence on $b$ is recovered 
 as a simple translation along the pressure axis. Returning to physical units, the dependence on both 
 $\bag$ and $m_s$ is recovered by appropriate scaling of that axis. 
   
\begin{figure}[htb]
\centering
\includegraphics[trim=0 0 0 0,clip,width=1\columnwidth]{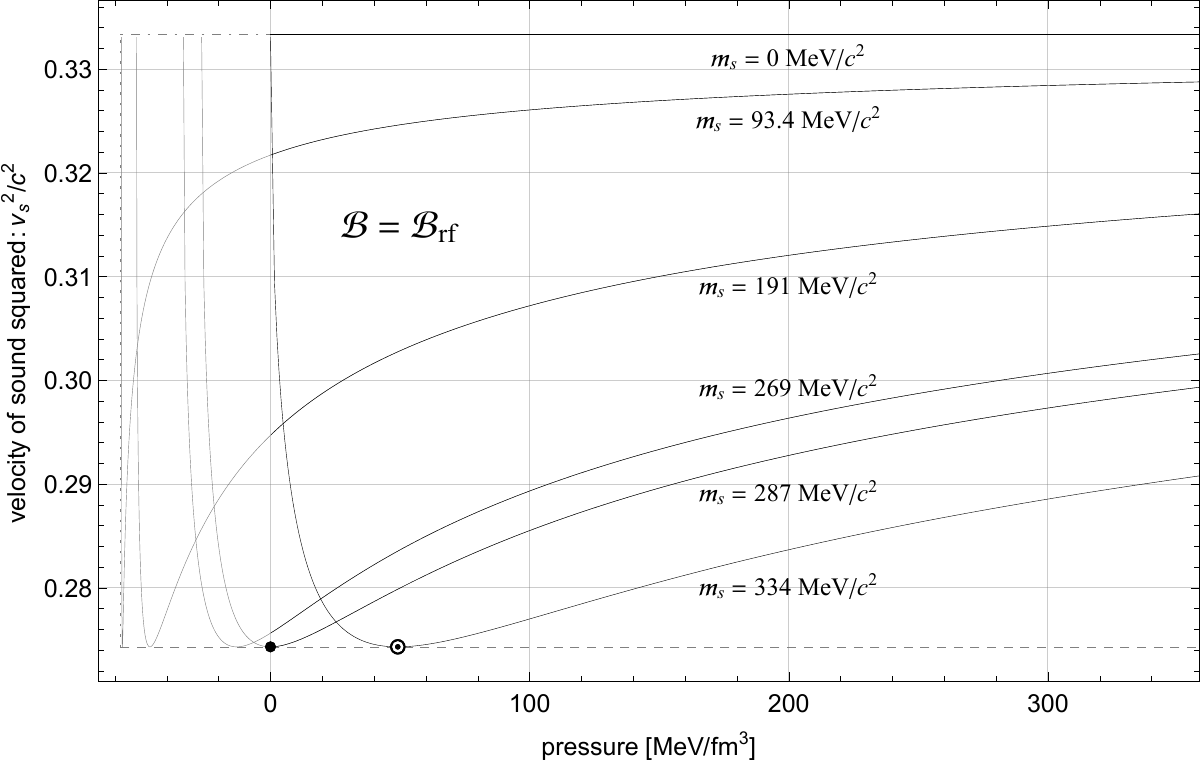}
\caption{\label{fig:sound-vel}The velocity of sound in SQM as a function of pressure 
for various s-quark masses $m_s$ at the reference bag constant $\bagref$. 
The empty center-dotted circle indicates the position 
of the velocity minimum for the highest allowable s-quark mass of ${\approx}334\,\MeV/c^2$. 
The filled black circle, located at vanishing pressure (the saturation point at the surface of the SQM star),
indicates the minimum velocity for the marginal s-quark mass of ${\approx}287\,\MeV/c^2$.
Above this mass, the speed of sound curves exhibit an inverted slope at lower pressures. The dashed line corresponds to the minimum velocity of sound 
$v_s{\approx} 0.5238 c$ in the model (independent of $\bagref$ and $m_s$). For $m_s=0$ the velocity is constant,  $v_s=c/\sqrt{3}$.   
}
\end{figure}

   The resulting speed of sound curve as 
 a function of pressure is shown for various  s-quark masses in \figref{fig:sound-vel}  at the reference bag constant $\bagref$. 
 It can be observed that, depending on the value of $b\propto\bag/m_s^4$ and the
  corresponding amount of translation of the curve, part of the curve falls below the physically interesting region of 
 $P\geqslant 0$ and `disappears from view'. At the same time, the profile becomes more compressed  as the s-quark mass  $m_s$ decreases.
 
 In the limit as $m_s\to0$, only the `conformal' tail $c_s=c/\sqrt{3}$ is visible, while in the opposite limit as $b\to b_m$, the entire
 curve enters the physical region $P>0$.
  For  sufficiently large $m_s$, above a certain threshold mass, the speed of sound curves exhibit an inverted slope at lower pressures compared to those with
 $m_s$ below this threshold. This indicates that
energy decreases less rapidly as pressure approaches zero toward the star's surface. For $m_s$ below the threshold mass, the rate of 
decrease  of energy density is significantly higher than for $m_s$ above the threshold. In other words,
for the same decrease in pressure, 
the energy density decreases less in the $m_s$--dominated regime than in the symmetric-like regime of low $m_s$.
 

\bibliographystyle{abbrvnat}
\bibliography{literatura} 

\end{document}